\documentclass{aa} 
\usepackage{graphicx}
\usepackage{pdflscape}
\usepackage{caption}
\usepackage{subcaption}
\usepackage[T1]{fontenc}
\begin{document} 
 
\title{A search for extended radio emission from 
selected compact galaxy groups}
 
\author { 
B{\l}azej Nikiel Wroczy\'nski\inst{1}
 \and M. Urbanik\inst{1}
 \and M. Soida\inst{1} 
\and R. Beck\inst{2}  
\and D.J. Bomans\inst{3} 
} 
\institute{Astronomical Observatory, Jagiellonian 
University, ul. Orla 171, PL 30-244 Krak\'ow, Poland 
\and Max-Planck-Institut f\"ur Radioastronomie, Auf dem 
H\"ugel 69, D53121 Bonn, Germany
\and Astronomisches Institut, Ruhr-Universit\"at-Bochum, 
D44780 Bochum, Germany } 
 
\offprints{B. Nikiel-Wroczy\'nski}
\mail{iwan@oa.uj.edu.pl}
\date{Received date/ Accepted date} 
 
\titlerunning{Radio emission from galaxy groups} 
\authorrunning{B. Nikiel -- Wroczy\'nski et al.}

\abstract 
{Studies on compact galaxy groups have led to the conclusion
that a plenitude of phenomena take place in between galaxies that
form them. However, radio data on these objects are extremely
scarce and not much is known concerning the existence and role of the
magnetic field in intergalactic space.} 
{We aim to study a small sample of galaxy groups that look 
promising as possible sources of intergalactic magnetic fields;
for example data from radio surveys suggest that most of the radio 
emission is due to extended, diffuse structures in and out of the 
galaxies.
}
{We used the Effelsberg 100 m radio telescope  at 4.85~GHz and NRAO VLA Sky Survey (NVSS) data at 1.40\,GHz.
After subtraction of compact sources we analysed the 
maps searching for diffuse, intergalactic radio emission.
Spectral index and magnetic field properties were derived.
}
{Intergalactic magnetic fields exist in groups
HCG\,15 and HCG\,60, whereas there are no signs of them in HCG\,68.
There are also hints of an intergalactic bridge in HCG\,44 at 4.85\,GHz.}
{Intergalactic magnetic fields exist in galaxy groups and their
energy density may be comparable to the thermal (X-ray) density,
suggesting an important role of the magnetic field in the intra-group
medium, wherever it is detected.}

\keywords{radio continuum:galaxies; galaxies: magnetic fields; galaxies:groups:individual: HCG15, HCG44, HCG60, HCG68; galaxies: intergalactic medium} 
 
\maketitle 
 
\section{Introduction \label{sec:intro}}

Hickson Compact Groups (HCG) constitute a scientifically 
interesting selection of dense galaxy systems containing a small number of galaxies. Originally
described by \citet{hickson82}, these objects are well known for the plenitude
of interaction-driven phenomena, such as morphological and kinematical peculiarities
of the member galaxies, bursts of star formation in their disks, nuclear radio and 
infrared emission due to starbursts, or active galactic nuclei (AGN; all reported by \citealt{hickson97}).
Also observed are giant gaseous outflows (e.g.  in HCG44, \citealt{serra13}), or even large-scale
shocks (e.g. in Stephan's Quintet; \citealt{appleton06}, \citealt{osullivan09}). 
Even though it has been proven that some of the originally defined HCGs
do not to fulfil the selection criteria, the acronym HCG stands
for a well-defined class of objects.\\

Among the different domains of the electromagnetic spectrum, the radio domain 
seems to be one of the least studied for the galaxy groups. This
is because in galaxy groups only compact sources of radio emission (e.g. the
aforementioned nuclei) can be observed at radio wavelengths reasonably
easily. Extended structures, with only a few exceptions, usually pose many
technical and observational difficulties.
Interferometric observations of objects in which bright compact sources are embedded 
in a weak diffuse emission are particularly exposed to the missing zero-spacing problem 
caused by incomplete spacing of the (u,v) plane. Reliable results need single-dish data.
However, in this case the beam-smearing effect mixes compact and diffuse emission. 
Additionally, a decaying electron
population, especially in the intergalactic areas, directs the attention
to lower frequencies.
However, for most of the radio telescopes such a combination results 
low angular resolution that is too low to study dense, and hence angularly small, galactic systems. 
Only in the advent of newest instruments, such as LOFAR \citep{LOFAR}, 
LWA \citep{LWA}, or MWA \citep{MWA} can high-resolution studies at low radio frequencies
be carried out. Unfortunately, galaxy group studies have not yet been performed
at any of these facilities.

Despite the impediments mentioned above, an increasing interest in studying the 
properties of the radio emission of HCGs can be observed recently. A selection
of compact groups has been studied using the GMRT \citep{giacintucci11} with
interesting results obtained. For example, Giacintucci et al. found the elliptical-dominated system
HCG\,15 has a large halo of radio emission that is visible both at 240 and 612\,MHz.
Another compact group, HCG\,62, was also featured in their study, and in that
case, extended emission was detected as well. A further example of a compact group that is
a radio emitter is the Stephan's Quintet, also known as HCG\,92. Moreover, the 
Quintet is, so far, the only compact group for which the polarised intergalactic
emission was detected \citep{bnw13B} and magnetic field properties \citep{xu03,bnw13B} 
were studied; the study of two southern compact galaxy groups \citep{farnes14} did
not reveal any intergalactic polarised structures.\\

In our paper on HCG\,92 we focused on the magnetic field properties. We found that the energy
density of the magnetic field in the shock region compares with that of the
thermal component 
(estimated using the X-ray data from \citealt{trinchold,trinchnew}, and \citealt{osullivan09}). 
Similar energy densities of the magnetic
field have also been reported for intergalactic structures in objects where no 
lookalikes of the shock region in the Quintet are present, for example in the Taffy Galaxies
\citep{condon93, condon02}. A strong suggestion is that
the magnetic field plays an important role in the dynamics of those groups where
it is present. Additionally, the orientation of the ordered magnetic field, which is detectable via the study of the polarised synchrotron emission, turns out to be a 
powerful tool for identifying possible intergalactic bridges and outflows.\\

In this paper we present the results of a pilot study of the magnetic field 
properties in a selected, small sample of HCGs. 
We used the 100 m Effelsberg radio telescope to observe selected Hickson groups for which archival high-resolution 
VLA data from the NRAO VLA Archive Survey (NVAS) showing no extended emission, but only compact structures were available.
To ensure full data compatibility and also the best
compromise of sensitivity and resolution of the Effelsberg radio telescope, we chose the 
frequency of 4.85~GHz. 
Because the single-dish data are now accompanied by short-time 
high-resolution observations showing only compact structures,  unwanted contribution (after convolution)
can be safely subtracted from the single-dish observations leaving only the extended emission.\\ 

Additionally, to obtain the spectral information we used the NRAO VLA Sky Survey
(NVSS; \citealt{NVSS}) data showing both compact and some diffuse emission at a resolution 
somewhat higher than that of the 100 m dish. At this frequency, the subtraction was 
also based on the archive data from the NVAS. Wherever possible, magnetic field estimations 
have been carried out and the morphology of the radio emission was investigated.

\section{Observations and data reduction}

\subsection{Selection criteria}

We have chosen four objects, for which the integrated flux density in the NRAO VLA Sky Survey (NVSS) at
1400\, MHz -- representing the emission of both extended and compact sources --
is significantly higher than the flux density in the high-resolution archive data,
representing the compact sources exclusively.The list consists of HCG\,15, 44, 60,\, and 68.\\

\begin{table}
\caption{\label{obsdata}Basic information on the datasets used in this study.}
\begin{center}
\begin{tabular}{rrrrr}
\hline
\hline
 HCG    & No. of cov.   & TOS$^{*}$     & r.m.s.        & r.m.s.        \\
 No.    &               & Effelsberg    &               & NVSS  \\
        &               & [h]           &  [mJy\slash beam]     &  [mJy\slash beam]   \\
\hline
 15     & 24            & 12.0          & 0.5           & 0.30          \\
 44     & 15            & 7.5           & 0.7           & 0.37          \\
 60     & 2             & 1.0           & 4.0           & 0.35          \\
 68     & 31            & 15.5          & 0.5           & 0.30          \\
\hline
\end{tabular}
\end{center}
\begin{flushleft}
$^{*}$ Time On Source
\end{flushleft}
\end{table}

The 4.85\,GHz radio data used in our study were recorded at the 100 m
radio telescope in Effelsberg using a dual-horn backend with an ability 
to detect Stokes I, Q, U, and V signals. The total bandwidth was 500\,MHz. For 
each of the targets, we obtained several coverages that were scanned in the
Az-El frame. All these coverages were reduced in the same manner via the NOD2 package \citep{nod2}; 
radio frequency interferences (RFI) and luminous sources detected by one 
horn only were flagged in the original (beam-switched) maps, which were
then restored, mirrored, and translated to the RA-Dec scheme. The total number
of coverages that went through this procedure as well as other basic 
information on the single-dish data can be found in Table~\ref{obsdata}.
For each of the groups, component 
maps were combined in the Fourier domain so the scanning effects can be reduced
\citep{emerson88}. The flux density scale, both for total power and linear polarisation,
was established using observations of 
3C\,48, 3C\,147, and 3C\,286, according to values given by 
\citet{mantovani09} (absolute calibration based on \citealt{baars}).
We used the Astronomical Image Processing System (\textsc{aips})  to alter the geometry of the
final images to match that of the optical images and to combine the Q and U maps
to produce polarisation intensity and angle distributions.
The NVSS data were acquired from the NRAO Postage Stamp Server, thus were
not re-calibrated; therefore, the full processing scheme can be found in \citet{NVSS}.

To include the calibration uncertainties, we introduced a 5\% uncertainty for compact
sources for all the channels (peak values). In case of extended sources, the final uncertainty is 
calculated as a quadratic sum of the calibration uncertainty and the map r.m.s. 
noise multiplied by the square root of the number of beams per a given
structure.

\subsection{Subtraction of the compact structures}

\begin{table*}
\caption{\label{nvasdata}Basic information on archive data used for subtraction}
\begin{center}
\begin{tabular}{cclccrr}
\hline
\hline
 HCG    & Band  & Project       & Conf.         & Resolution$^*$        & Sensitivity$^*$ & Max. obs. str$^{**}$  \\
 No.    &       &               &               & [arcsec]              & [$\mu$Jy/beam]  & [arcsec]              \\
\hline
 15     & L     & AM108         & B             &  4.6$\times$3.6       & 11                      & 60                    \\
        & C     & AY171         & D             & 16.9$\times$13.2      & 11                      & 240                   \\
 44     & L     & AM24          & C             & 14.1$\times$12.4      & 19                      & 485                   \\
        & C     & AM188         & CD            & 12.1$\times$4.1\,\,\, & 6                       & 120                   \\
 60     & L     & AM108         & B             &  3.9$\times$3.3       & 8                       & 60                    \\
        & C     & AM188         & CD            & 11.0$\times$4.8\,\,\, & 8                       & 120                   \\
 68     & L     & MENO          & C             & 13.8$\times$10.5      & 30                      & 485                   \\
        & C     & AW137         & C             &  4.1$\times$3.5       & 6                       & 120                   \\
\hline
\end{tabular}
\end{center}
\begin{flushleft}
 $^{*}$ Original, before convolution\\
 $^{**}$ Maximal observable structure; length of the observations taken into account
\end{flushleft}

\end{table*}

The beam of the single-dish observations is large enough so that unwanted contribution
from compact sources, i.e. background objects, disks of member galaxies, or AGN activity within the groups, can be 
smeared together with the extended intergalactic emission, which is the target of our 
study. Before the magnetic field can be analysed, subtraction of such sources must be
performed. We searched the NRAO Archive and the NVAS for
high-resolution observations of our target groups taken at 1.40 and 4.85\,GHz. We completed
a set of eight total power datasets, i.e. two per each group for 4.85 and 1.40\,GHz maps, which  represented the compact
sources contribution. These datasets (details on them can be found in Table~\ref{nvasdata}) were reduced following the standard VLA continuum data reduction
procedure (as outlined  in the \textsc{aips} cookbook). At 4.85\,GHz, these maps were then convolved with the Effelsberg beam image
(acquired from the calibrator observations) using \textsc{aips}. As
a consistency check of the flux density scale, we performed a similar operation with a Gaussian image
of the same half power beam width (HPBW) and then with (u,v)-tapered maps. In both cases, the resulting images were consistent with those obtained using the first
method within the derived flux density uncertainties. Convolved VLA maps were subtracted from 
the Effelsberg observations. 
A similar procedure was applied to the 1.40\,GHz data with the exception that the beam was assumed to have a Gaussian
profile, and the beam shape was assumed to be same as for the NVSS (circular, HPBW of 45 arcseconds).

\begin{sidewaysfigure*}
    \begin{subfigure}[p]{0.33\textwidth}
    \includegraphics[width=\textwidth]{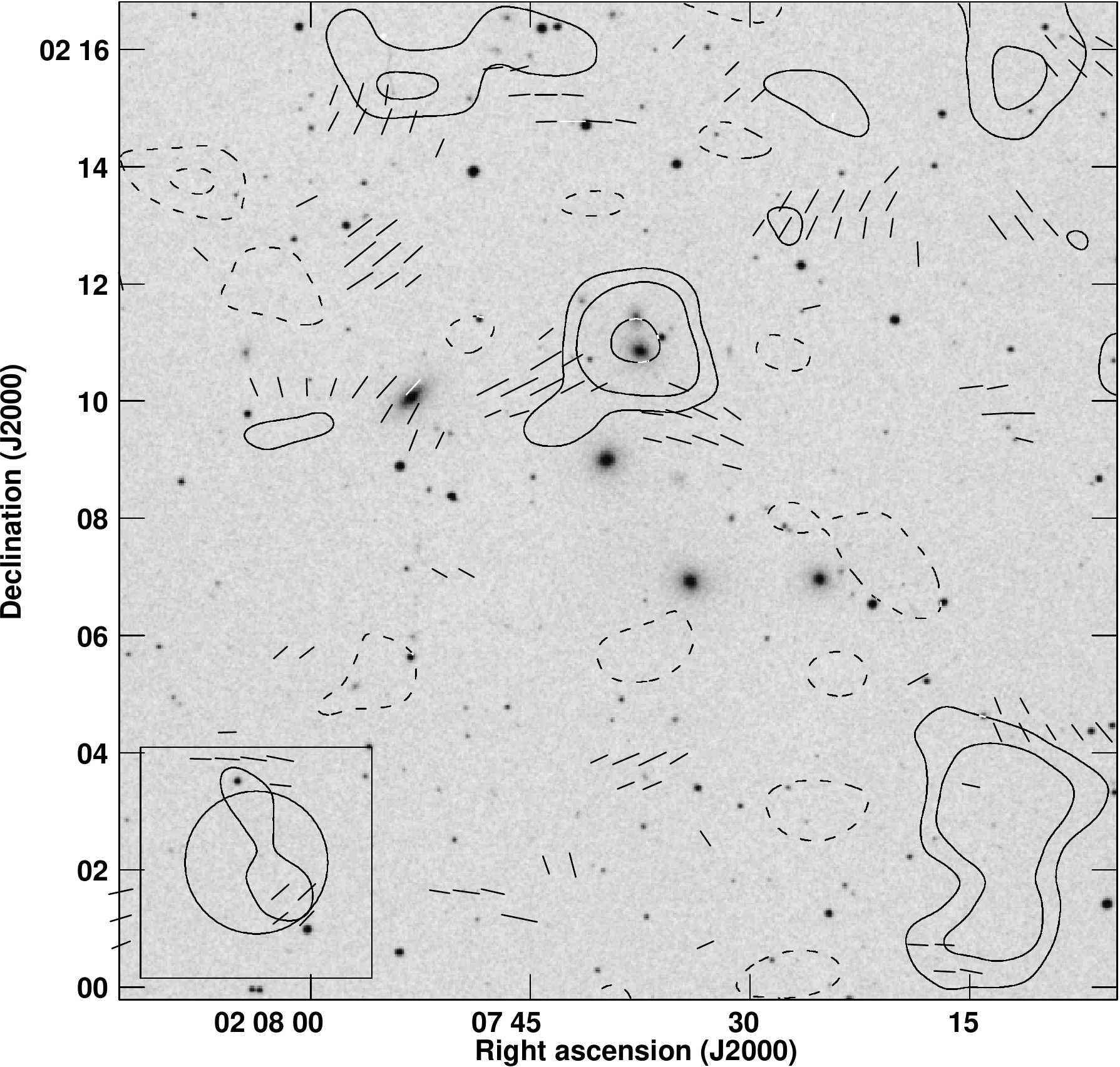}
    \caption{}
    \label{15eff}
    \end{subfigure}%
    \begin{subfigure}[p]{0.33\textwidth}
    \includegraphics[width=\textwidth]{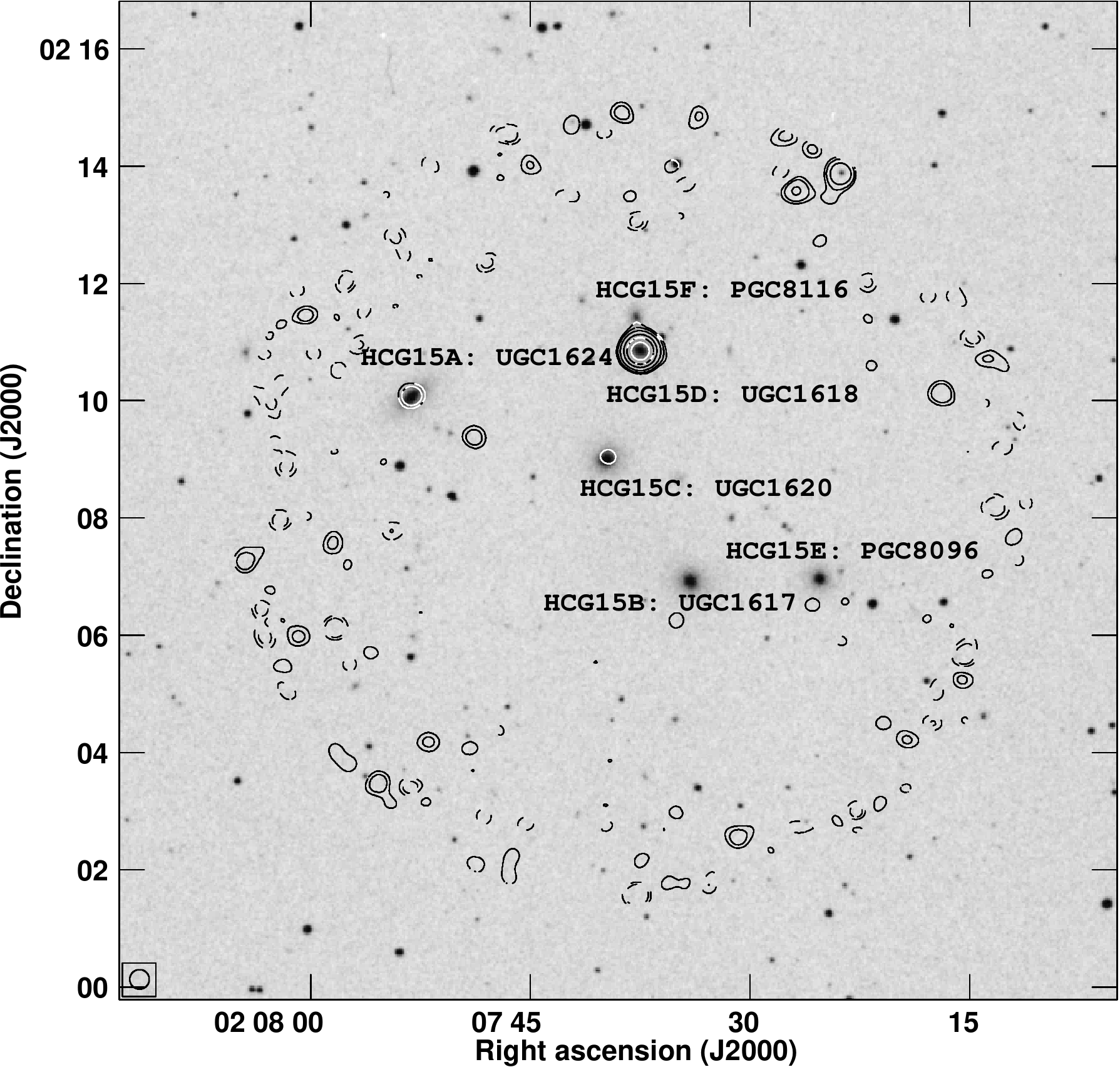}
    \caption{}
    \label{15cnvas20}
    \end{subfigure}%
    \begin{subfigure}[p]{0.33\textwidth}
    \includegraphics[width=\textwidth]{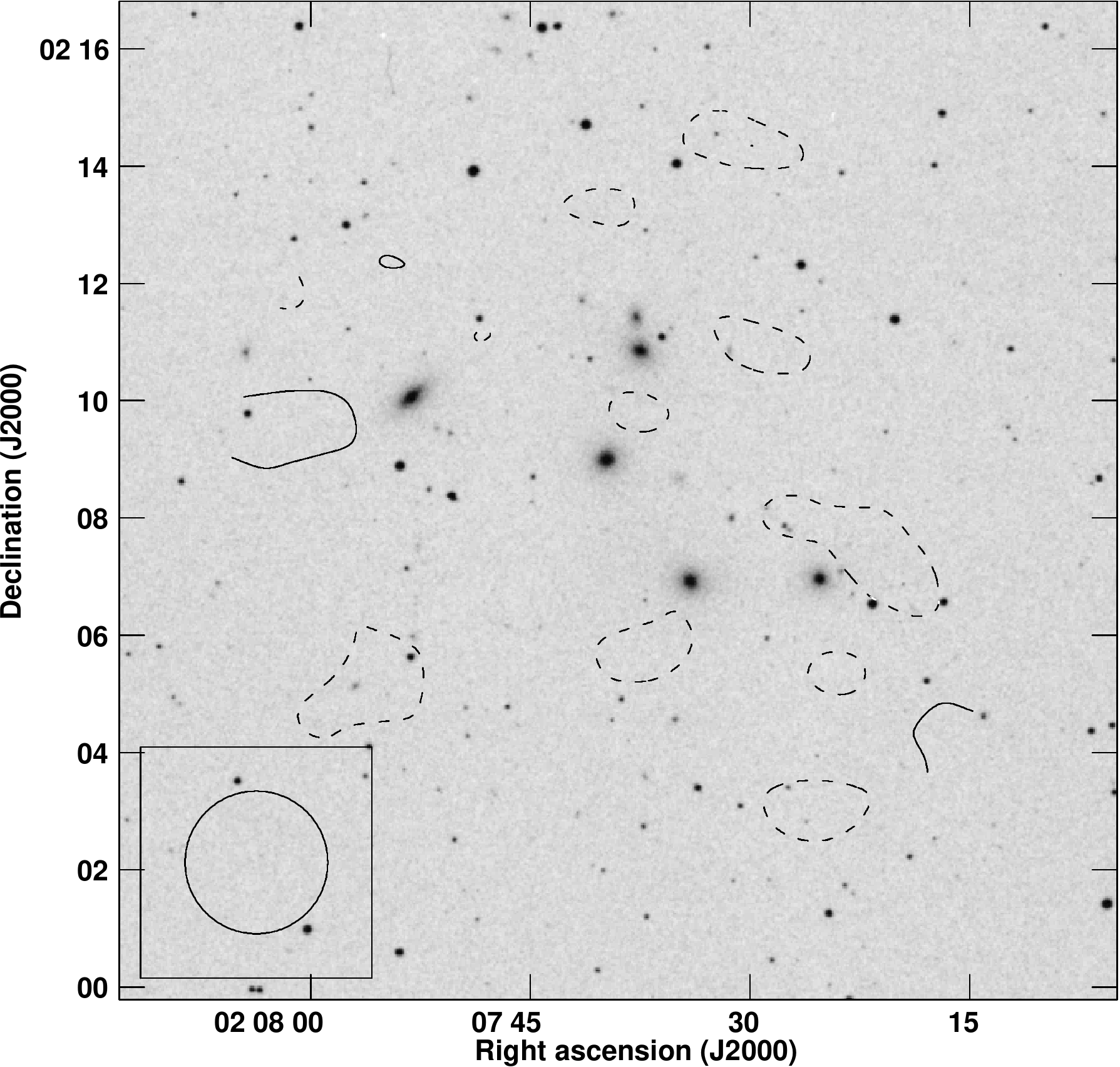}
    \captionsetup{width=0.9\textwidth}
    \caption{}
    \label{15esub}
    \end{subfigure}%

    \begin{subfigure}[p]{0.33\textwidth}
    \includegraphics[width=\textwidth]{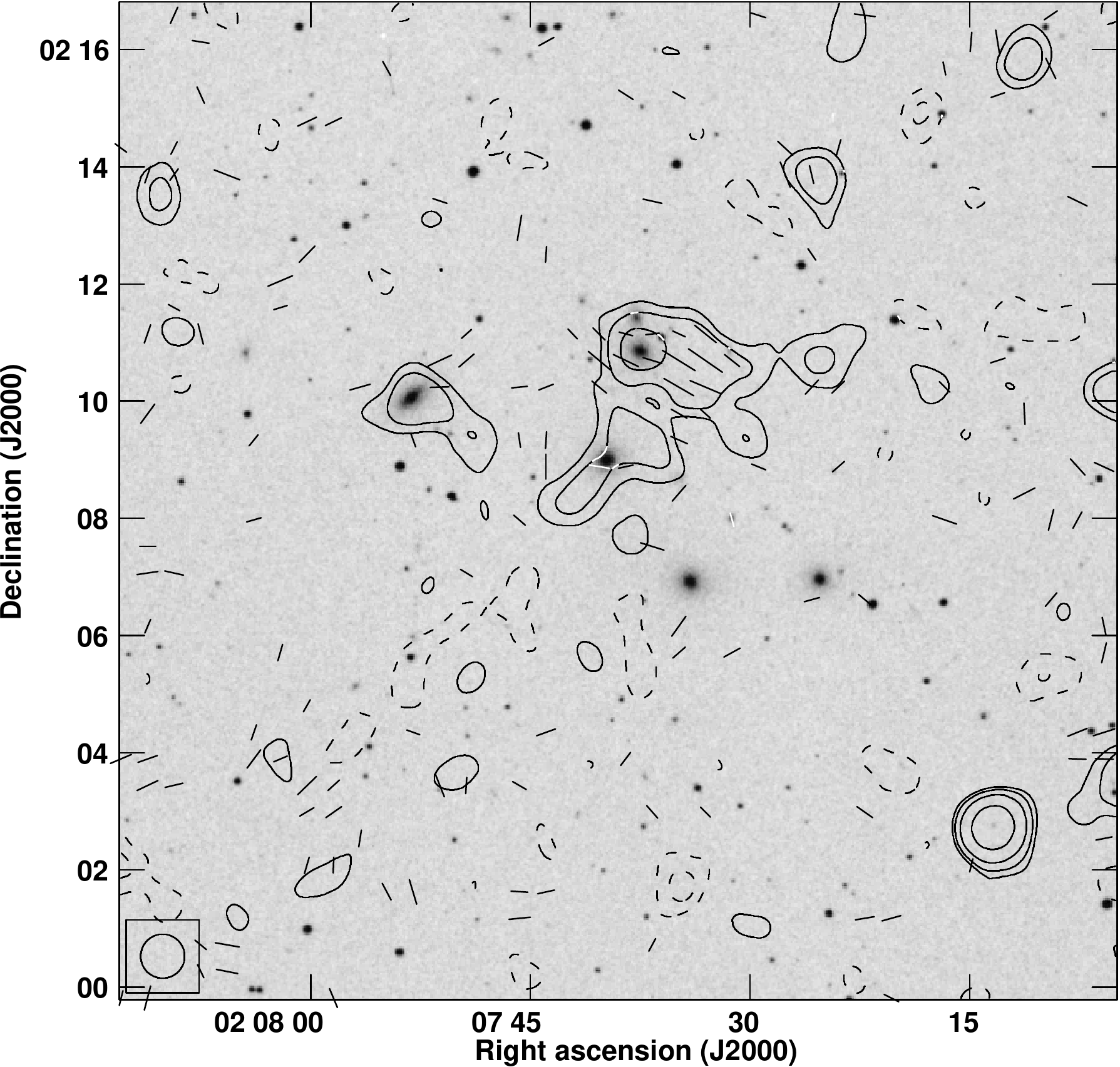}
    \captionsetup{width=0.9\textwidth}
    \caption{}
    \label{15nvss}
    \end{subfigure}%
    \begin{subfigure}[p]{0.33\textwidth}
    \includegraphics[width=\textwidth]{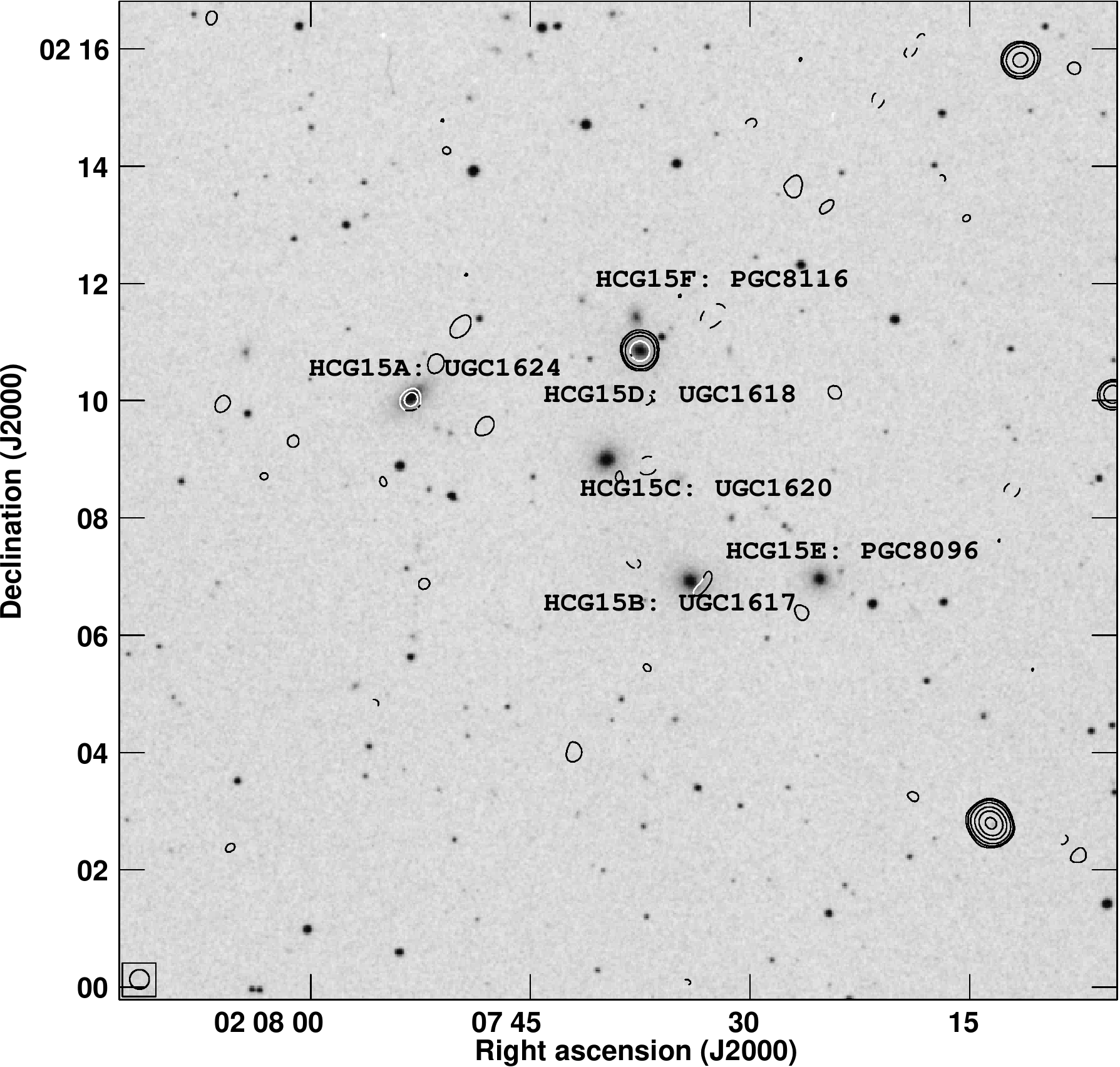}
    \caption{}
    \label{15lnvas20}
    \end{subfigure}%
    \begin{subfigure}[p]{0.33\textwidth}
    \includegraphics[width=\textwidth]{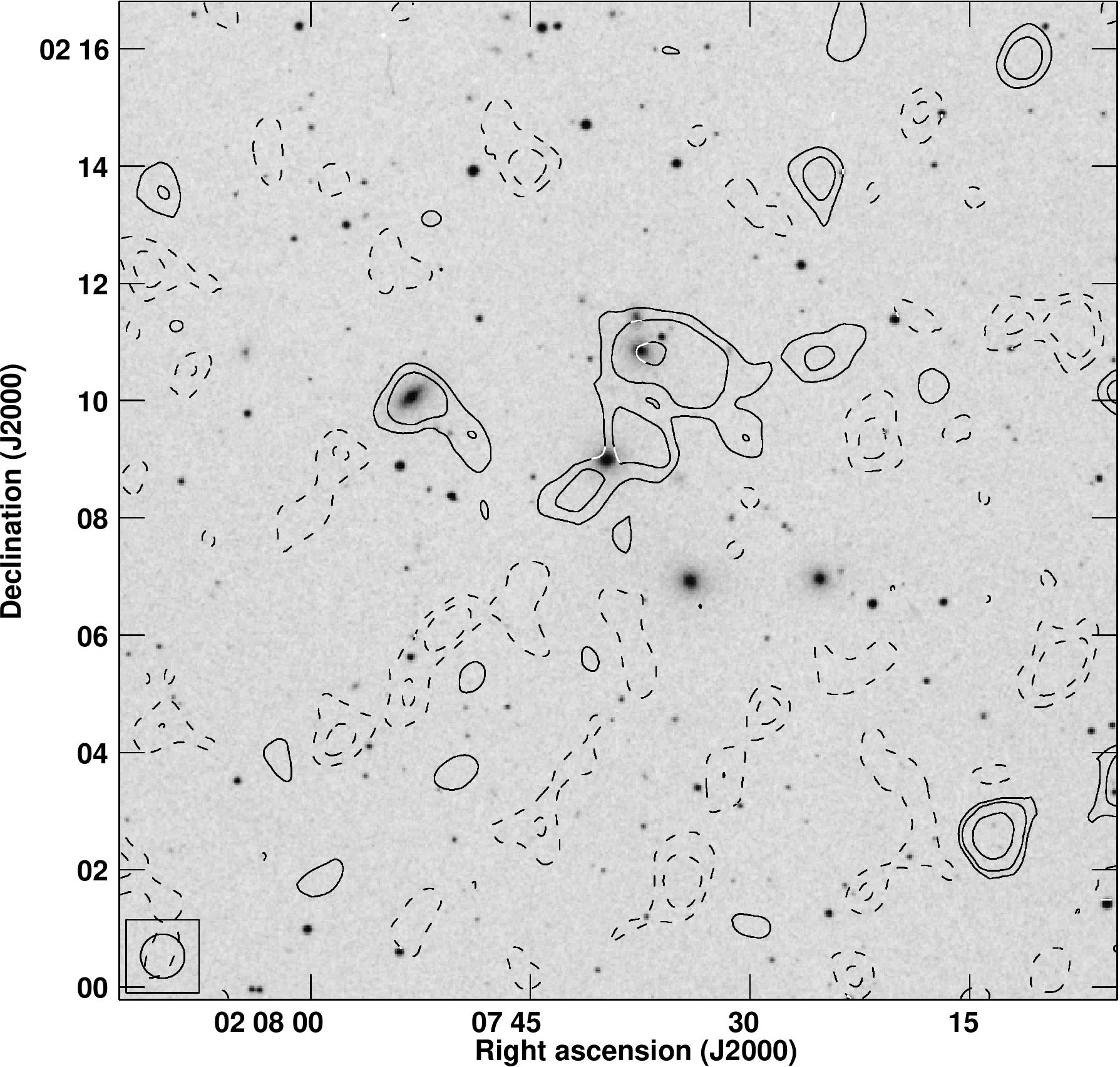}
    \captionsetup{width=0.9\textwidth}
    \caption{}
      \label{15nvsub}
     \end{subfigure}%
    \caption{
          Radio maps of HCG\,15. \textbf{Upper panel:} Effelsberg maps at 4.85\,GHz are shown. 
          \textbf{Lower panel:} NVSS maps at 1.40\,GHz are shown. \textbf{Common for both frequencies:}
          The left map is the TP emission with apparent B vectors overlaid.
          The central map is the TP emission from the NVAS data smoothed to 20 arcsec resolution.
          The right map is the TP emission with compact sources subtracted.
          The background map is a POSS-II R-band image.
          The contour levels are $-5,-3$(dashed), $3,5,10 \times$ r.m.s. noise level. 
          The beam is represented by a circle in the lower left corner of the image.
          The 1 arcsec length of the apparent B vectors corresponds to 0.03 mJy/beam.
          1 arcminute is equal to $\approx$ 27\,kpc at the position of HCG\,15.
          }
     \label{fig15}
\end{sidewaysfigure*}

\begin{sidewaysfigure*}
    \begin{subfigure}[p]{0.33\textwidth}
    \includegraphics[width=\textwidth]{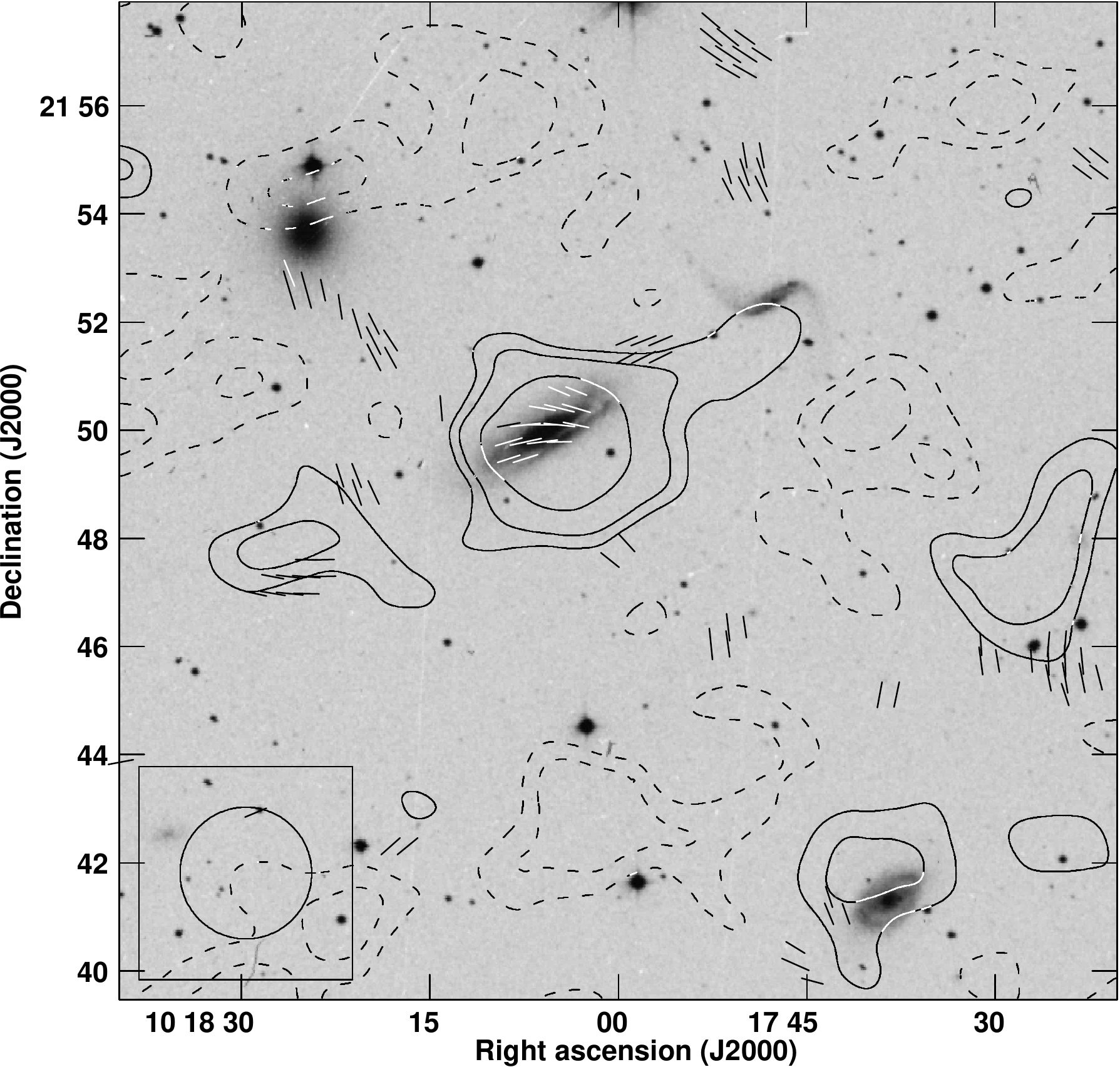}
    \caption{}
    \label{44eff}
    \end{subfigure}%
    \begin{subfigure}[p]{0.33\textwidth}
    \includegraphics[width=\textwidth]{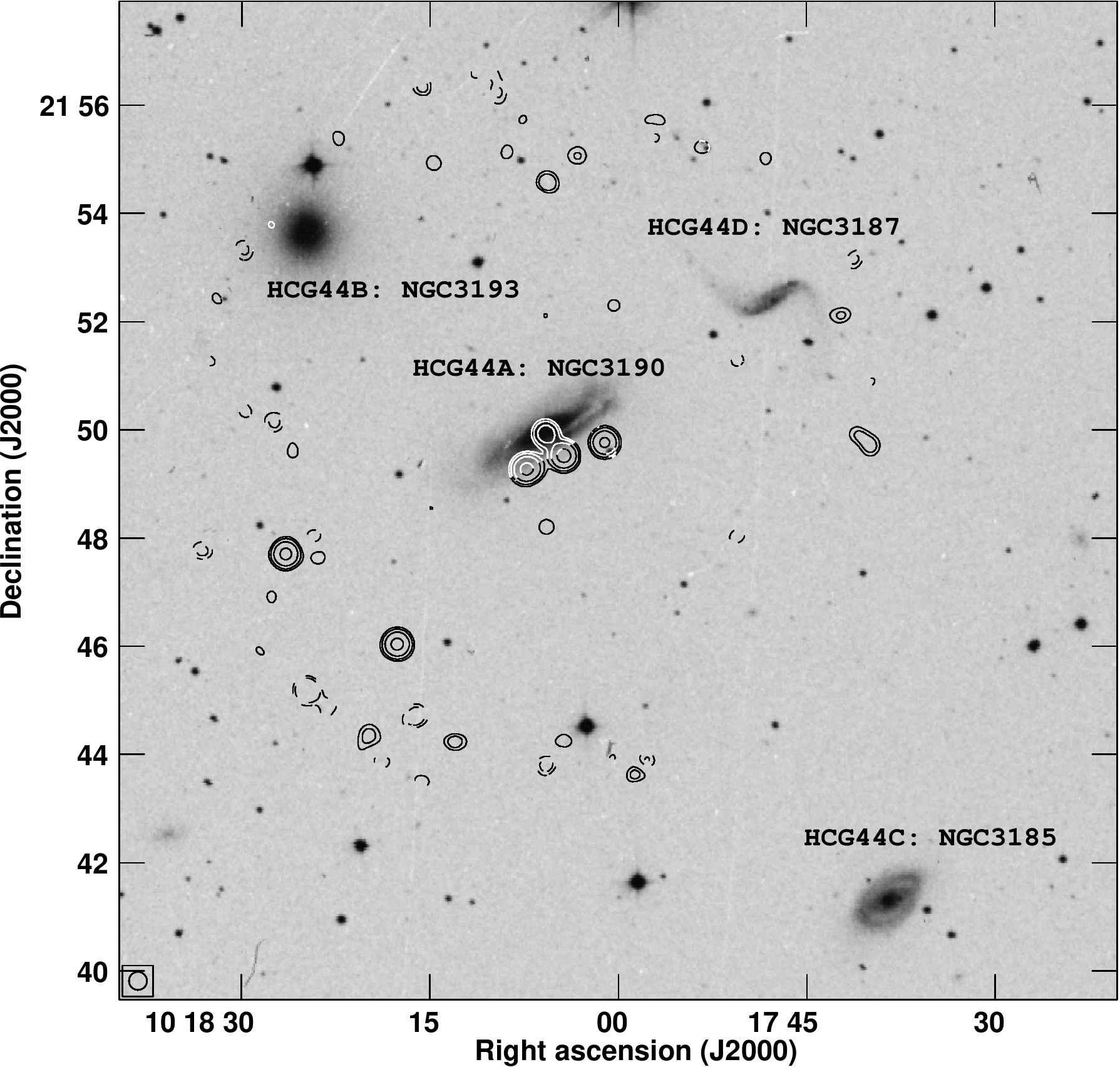}
    \caption{}
    \label{44cnvas20}
    \end{subfigure}%
    \begin{subfigure}[p]{0.33\textwidth}
    \includegraphics[width=\textwidth]{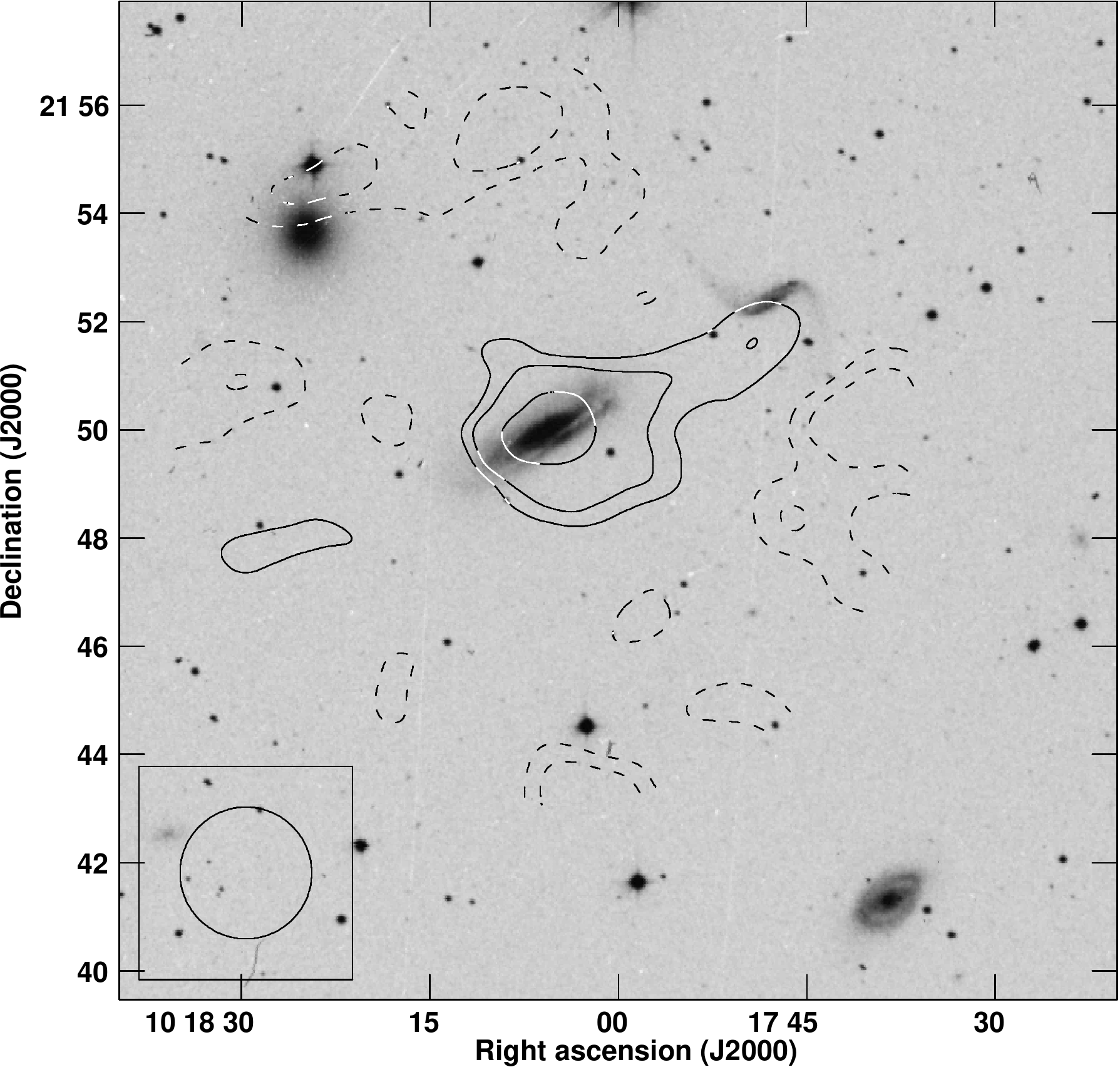}
    \caption{}
    \label{44esub}
    \end{subfigure}%

    \begin{subfigure}[p]{0.33\textwidth}
    \includegraphics[width=\textwidth]{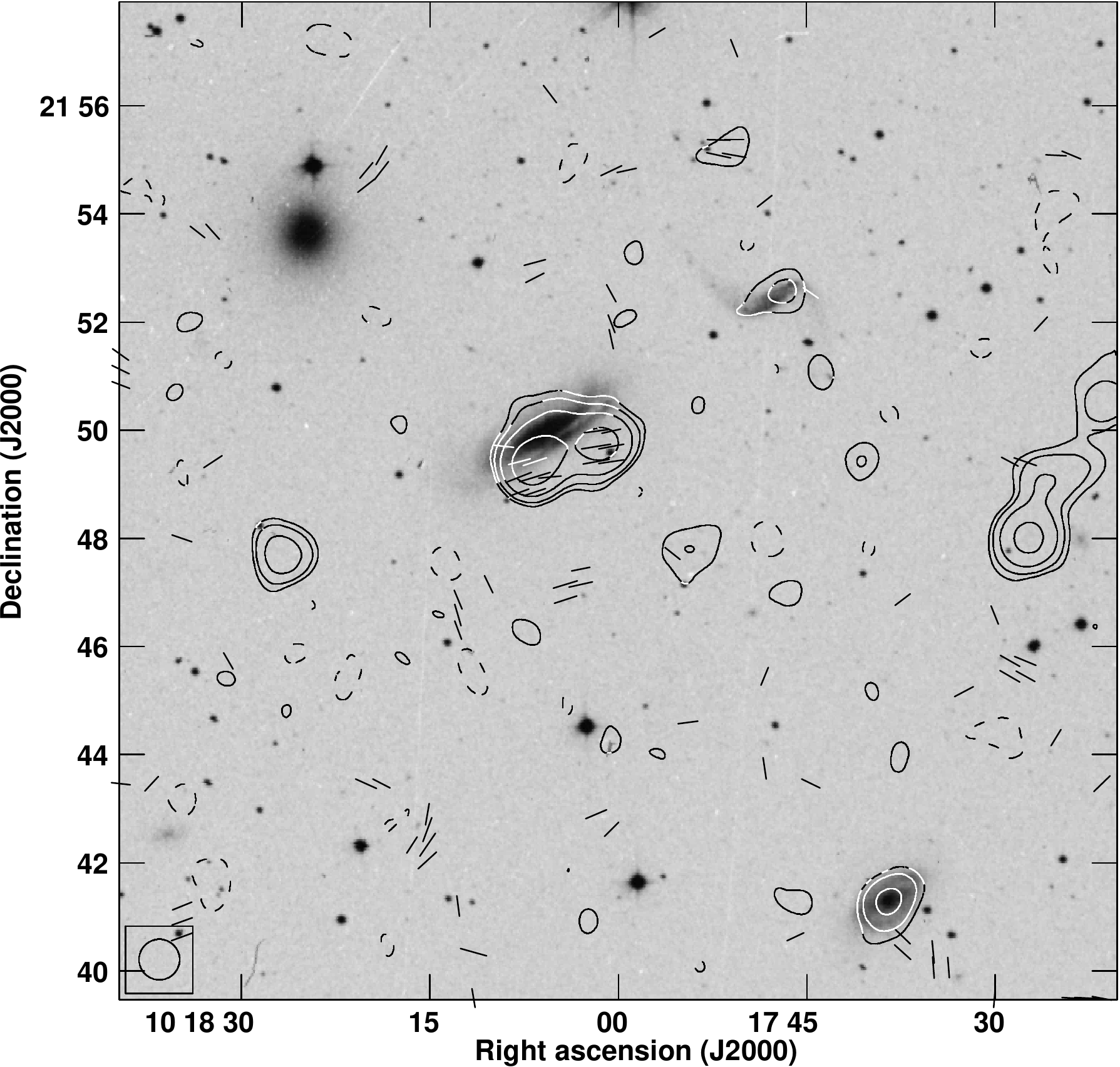}
    \captionsetup{width=0.9\textwidth}
    \caption{}
    \label{44nvss}
    \end{subfigure}%
    \begin{subfigure}[p]{0.33\textwidth}
    \includegraphics[width=\textwidth]{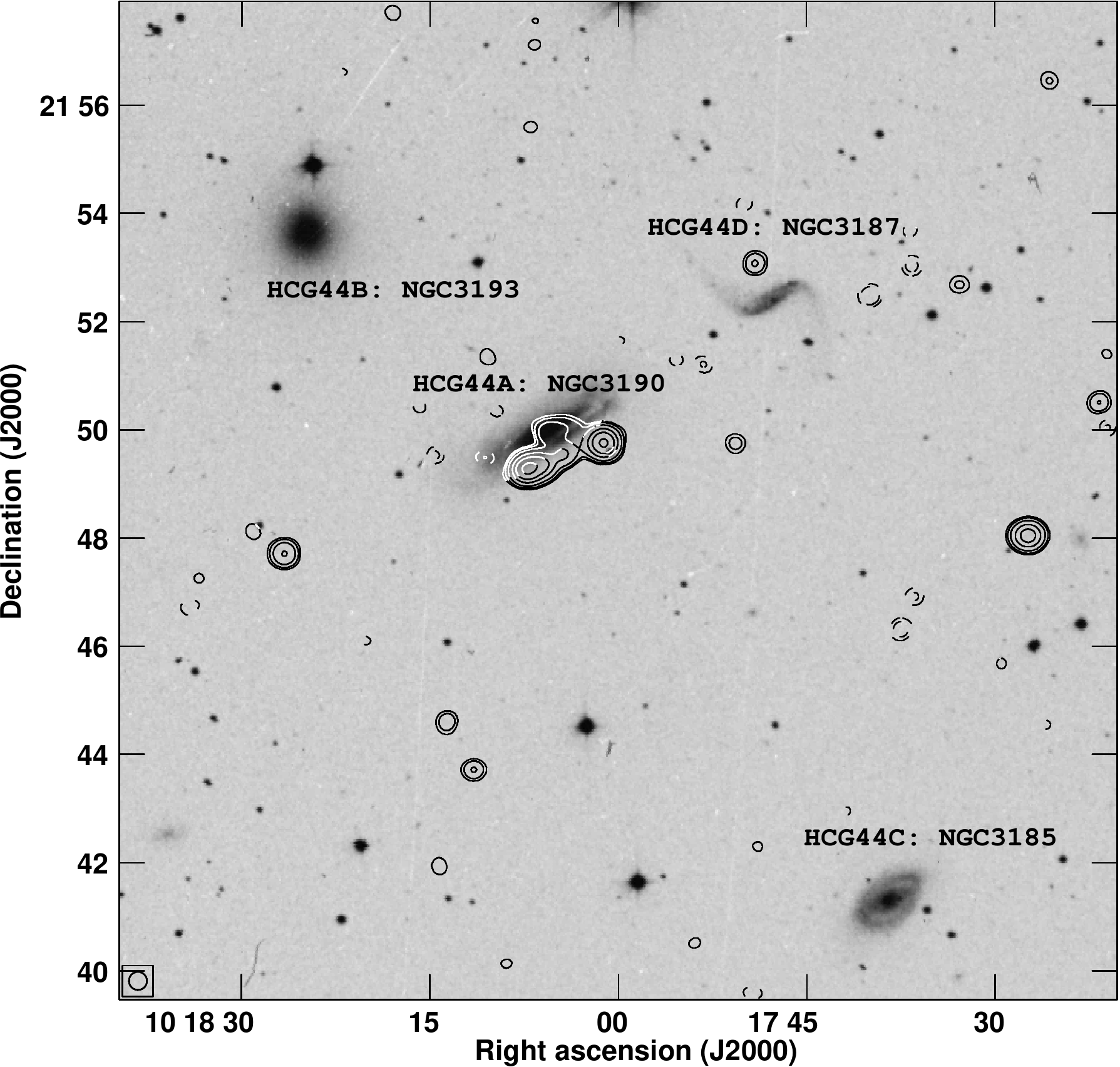}
    \caption{}
    \label{44lnvas20}
    \end{subfigure}%
    \begin{subfigure}[p]{0.33\textwidth}
    \includegraphics[width=\textwidth]{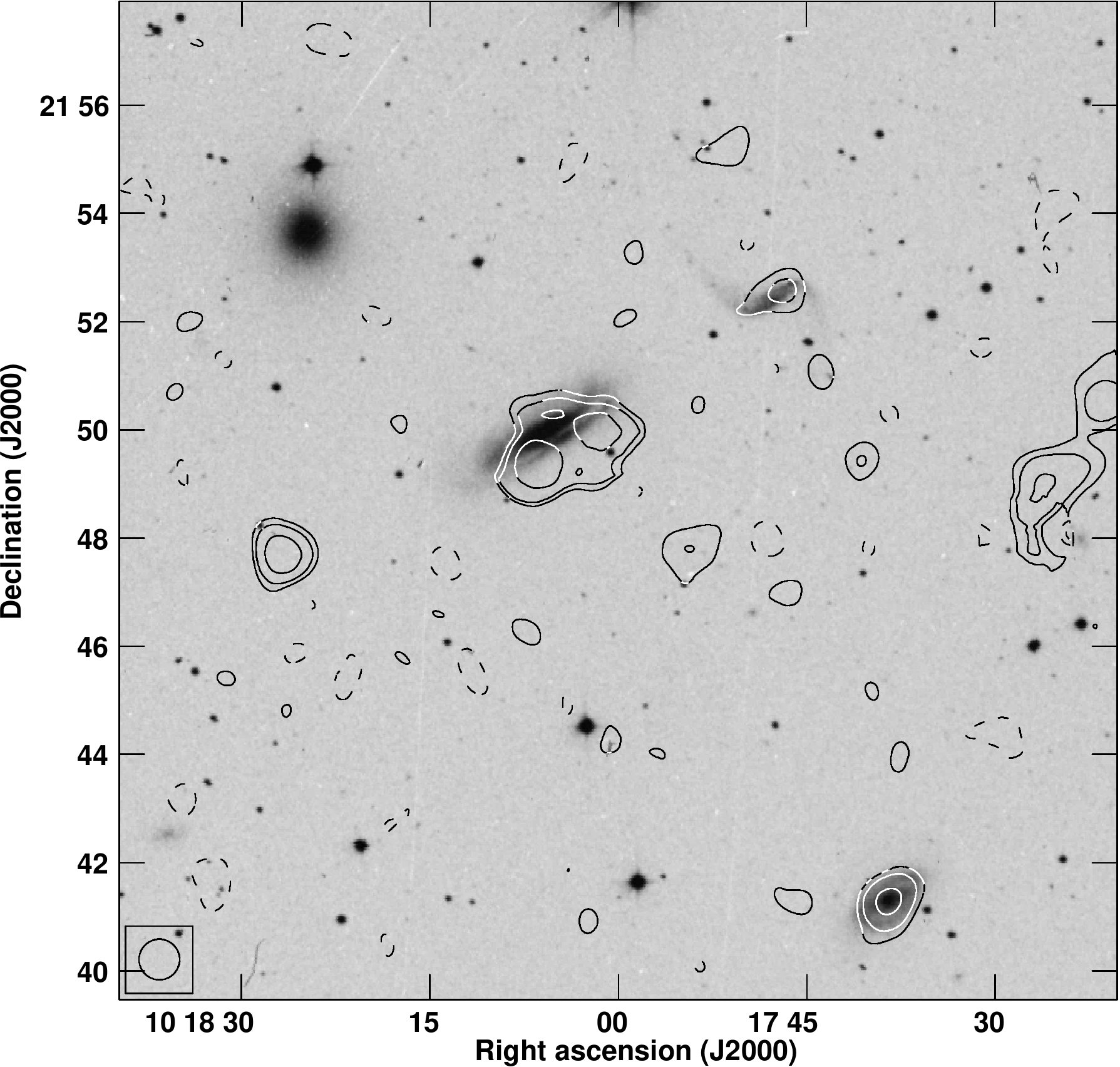}
    \caption{}
      \label{44nvsub}
     \end{subfigure}%
    \caption{
          Radio maps of HCG\,44. \textbf{Upper panel:} Effelsberg maps at 4.85\,GHz are shown. 
          \textbf{Lower panel:} NVSS maps at 1.40\,GHz are shown. \textbf{Common for both frequencies:}
          The left map is the TP emission with apparent B vectors overlaid.
          The central map is the TP emission from the NVAS data smoothed to 20 arcsec resolution.
          The right map is the TP emission with compact sources subtracted.
          The background map is a POSS-II R-band image.
          The contour levels are $-5,-3$(dashed), $3,5,10,25 \times$ r.m.s. noise level. 
          The beam is represented by a circle in the lower left corner of the image.
          The 1-arcsec length of the apparent B vectors corresponds to 0.03 mJy/beam.
          1 arcminute is equal to $\approx$ 7\,kpc at the position of HCG\,44.
        }
    \label{fig44}
\end{sidewaysfigure*}

\begin{sidewaysfigure*}
    \begin{subfigure}[p]{0.33\textwidth}
    \includegraphics[width=\textwidth]{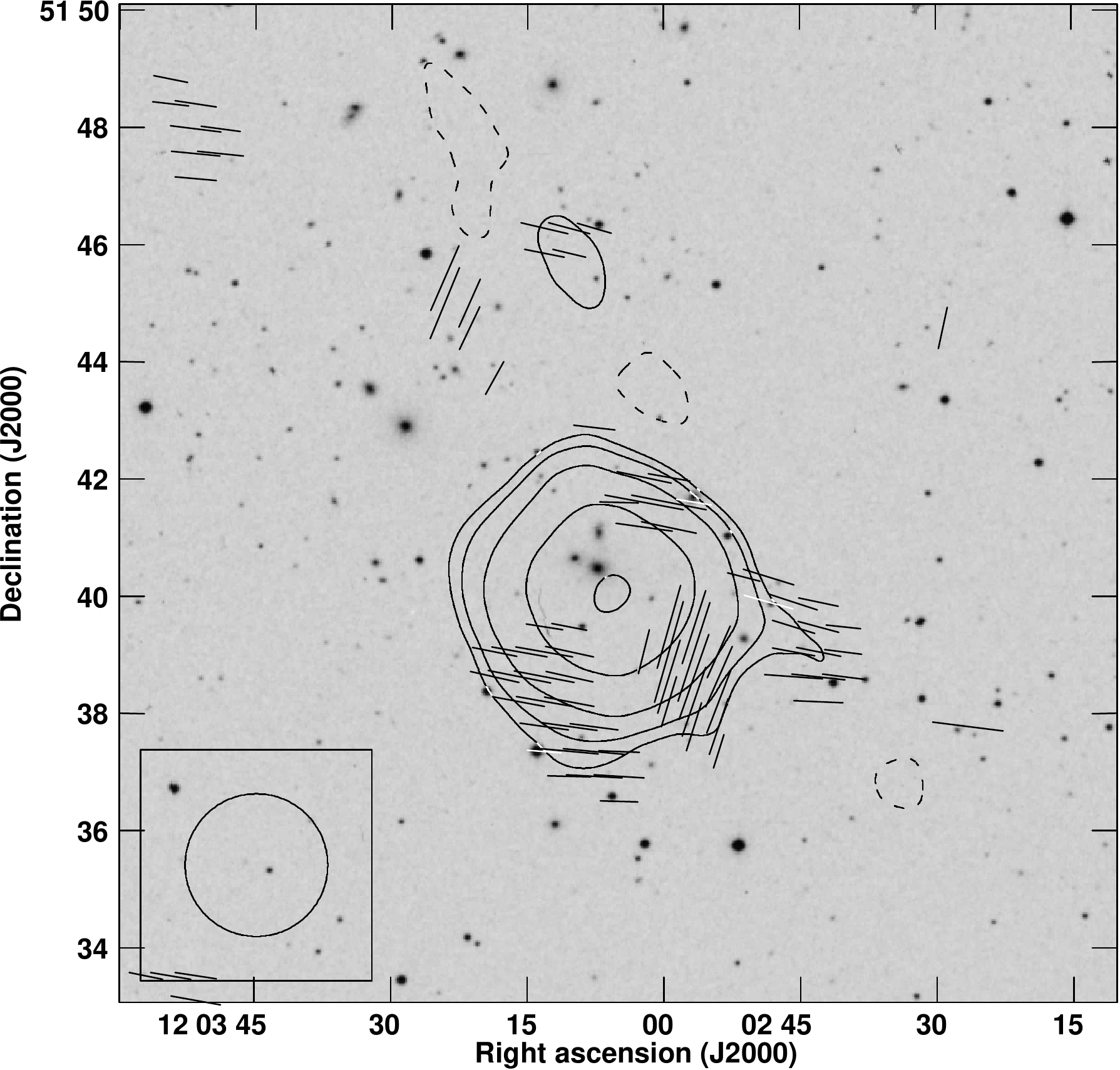}
    \caption{}
    \label{60eff}
    \end{subfigure}%
    \begin{subfigure}[p]{0.33\textwidth}
    \includegraphics[width=\textwidth]{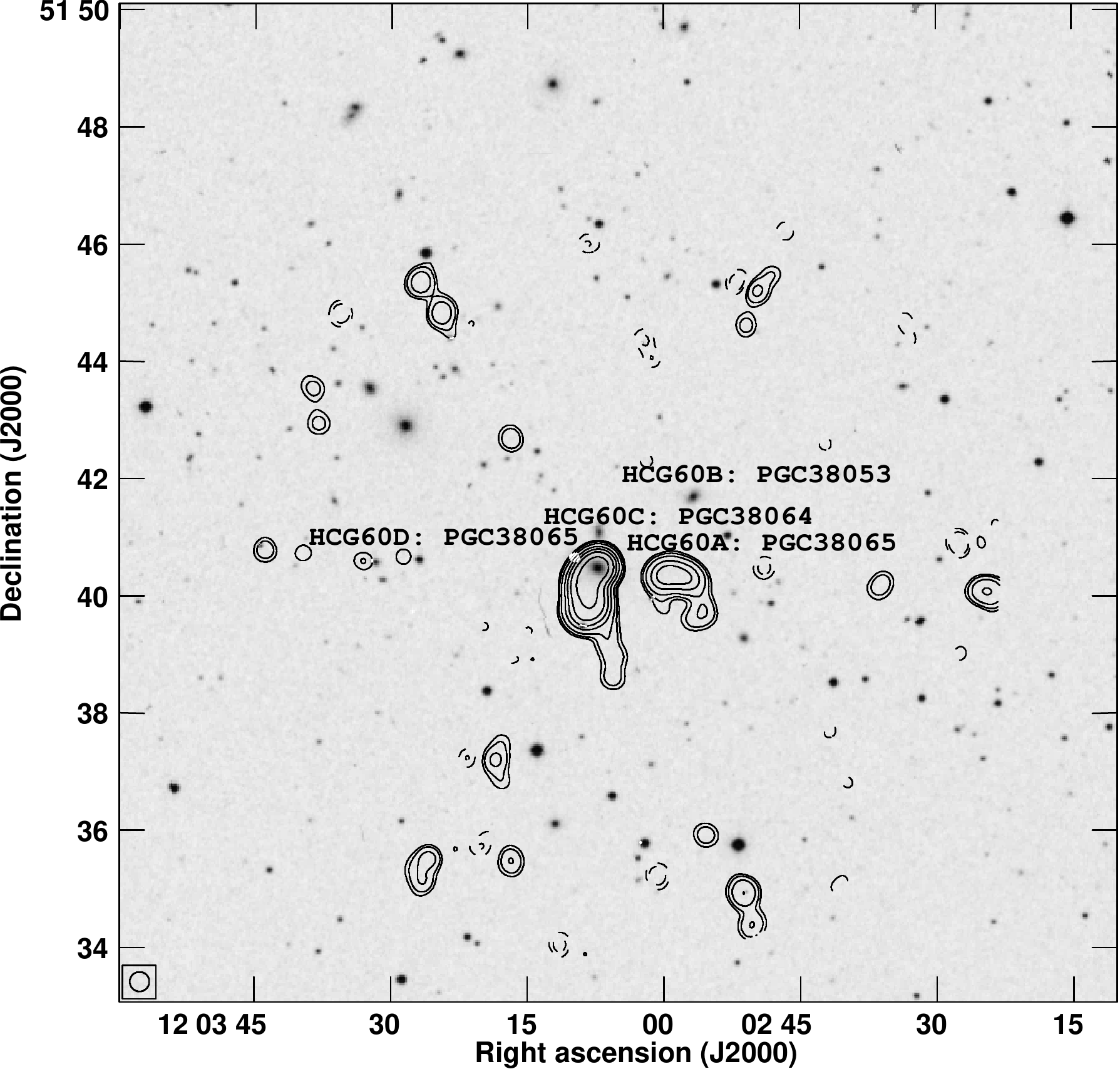}
    \caption{}
    \label{60cnvas20}
    \end{subfigure}%
    \begin{subfigure}[p]{0.33\textwidth}
    \includegraphics[width=\textwidth]{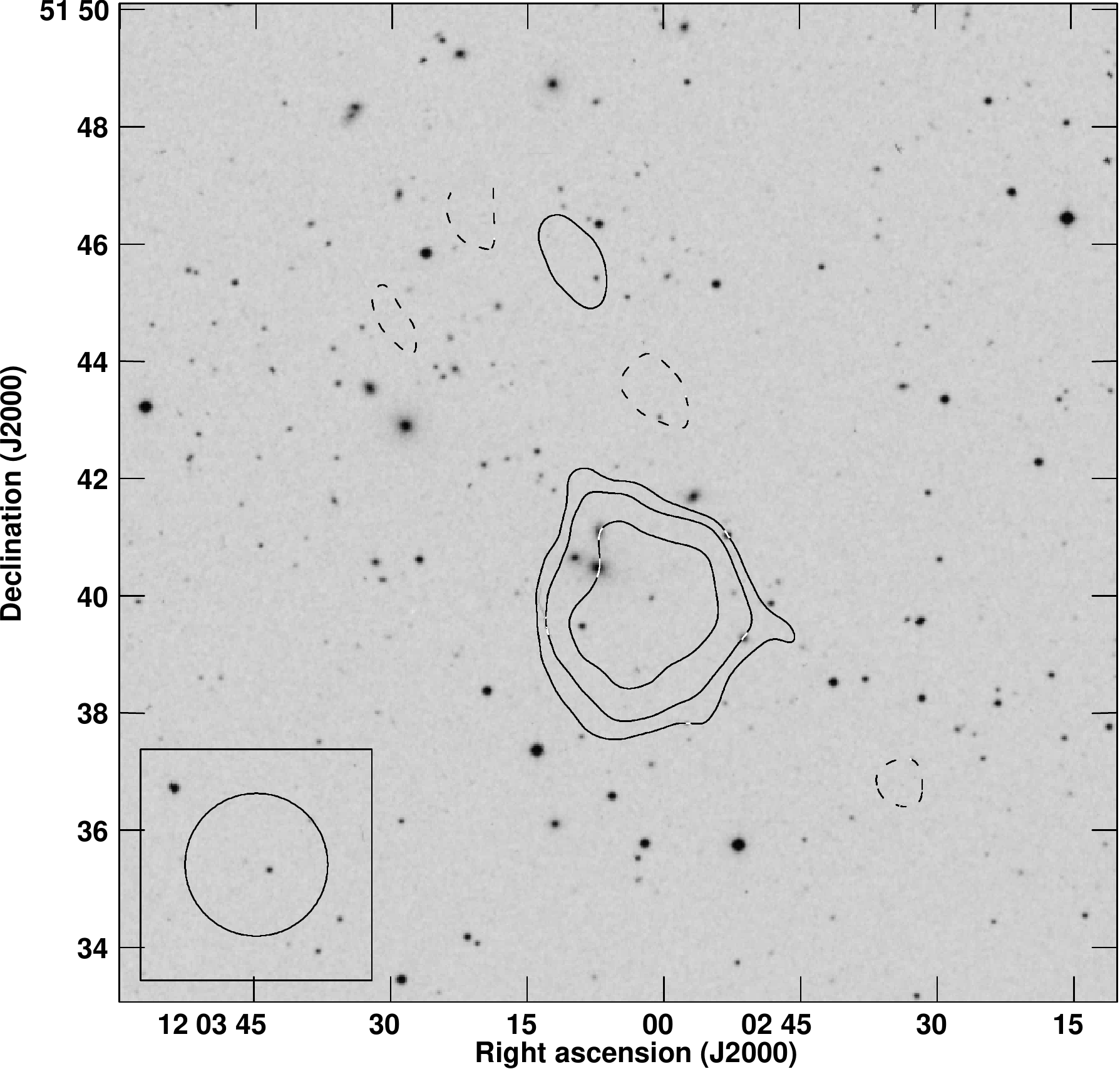}
    \captionsetup{width=0.9\textwidth}
    \caption{}
    \label{60esub}
    \end{subfigure}%

    \begin{subfigure}[p]{0.33\textwidth}
    \includegraphics[width=\textwidth]{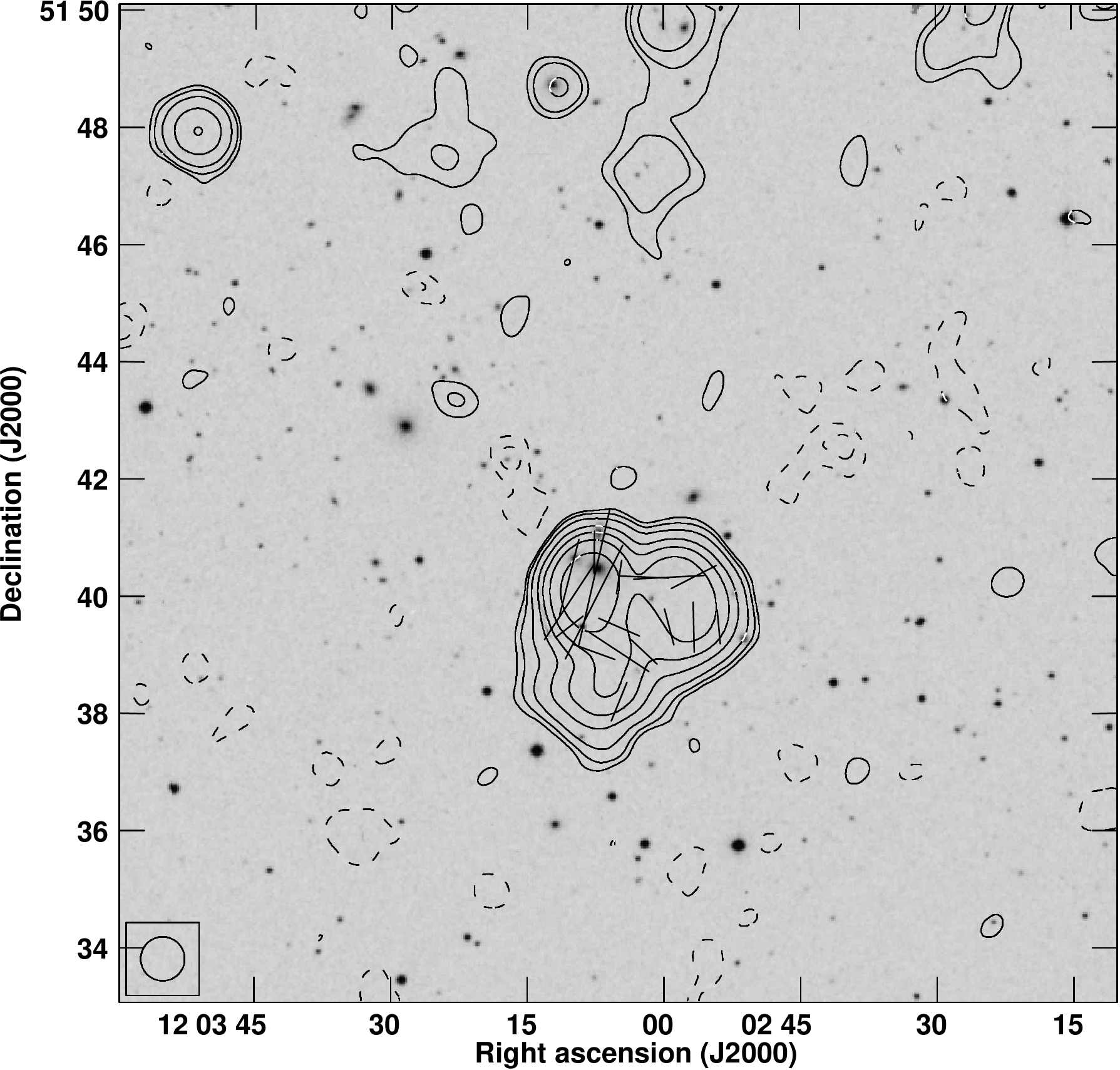}
    \captionsetup{width=0.9\textwidth}
    \caption{}
    \label{60nvss}
    \end{subfigure}%
    \begin{subfigure}[p]{0.33\textwidth}
    \includegraphics[width=\textwidth]{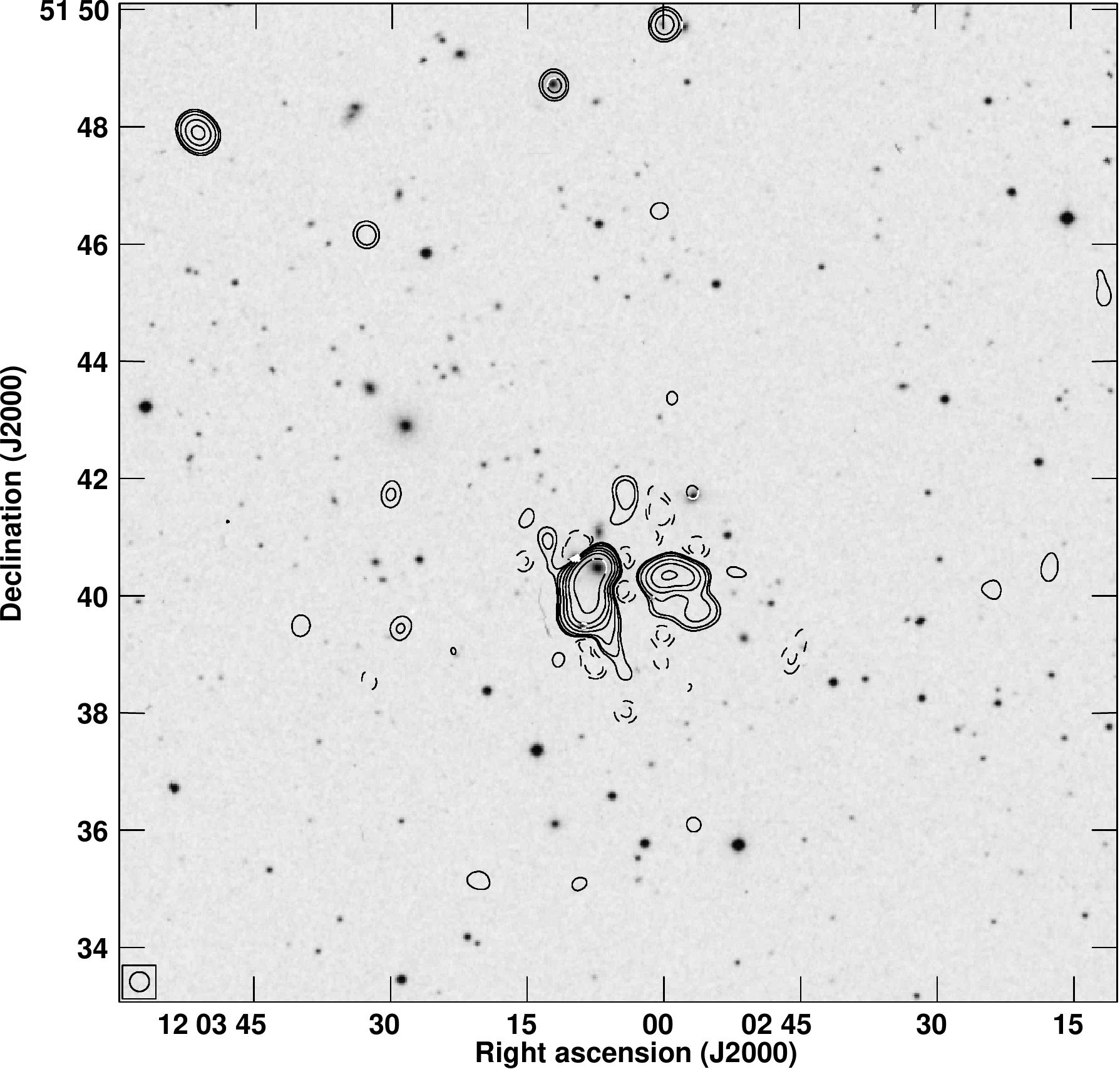}
    \caption{}
    \label{60lnvas20}
    \end{subfigure}%
    \begin{subfigure}[p]{0.33\textwidth}
    \includegraphics[width=\textwidth]{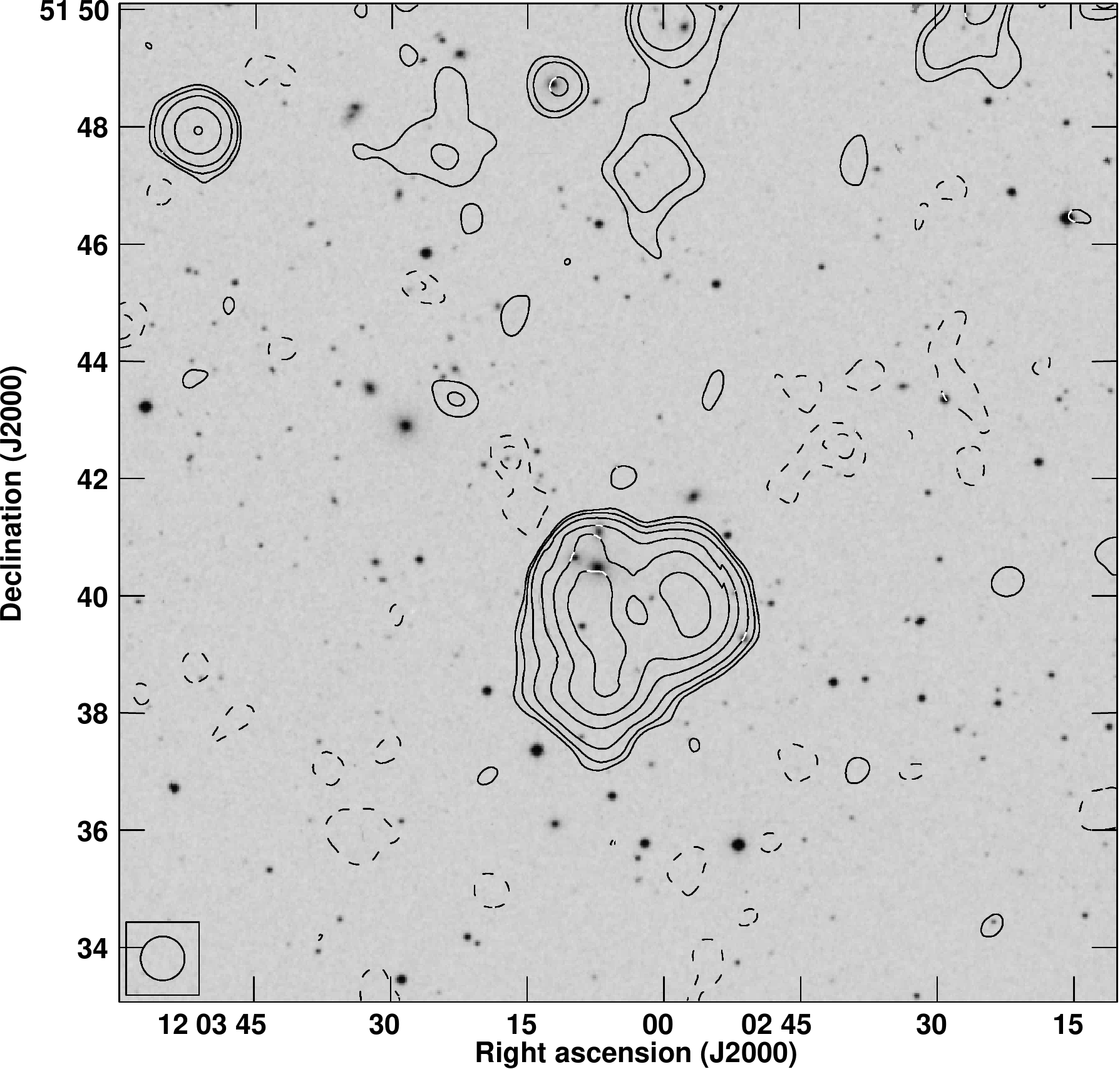}
    \captionsetup{width=0.9\textwidth}
    \caption{}
      \label{60nvsub}
     \end{subfigure}%
    \caption{
          Radio maps of HCG\,60. \textbf{Upper panel:} Effelsberg maps at 4.85\,GHz are shown. 
          \textbf{Lower panel:} NVSS maps at 1.40\,GHz are shown. \textbf{Common for both frequencies:}
          The left map is the TP emission with apparent B-vectors overlaid.
          The central map is the TP emission from the NVAS data, smoothed to 20 arcsec resolution.
          The right map is the TP emission with compact sources subtracted.
          The background map is a POSS-II R-band image.
          The background map is a POSS-II R-band image.
          The contour levels are $-5,-3$(dashed), $3,5,10,25,50,100,250\times$ r.m.s. noise level. 
          The beam is represented by a circle in the lower left corner of the image.
          The 1-arcsec length of the apparent B vectors corresponds to 0.06 mJy/beam;
          1 arcminute is equal to $\approx$ 75\,kpc at the position of HCG\,60.
        }
    \label{fig60}
\end{sidewaysfigure*}

\begin{sidewaysfigure*}
    \begin{subfigure}[p]{0.33\textwidth}
    \includegraphics[width=\textwidth]{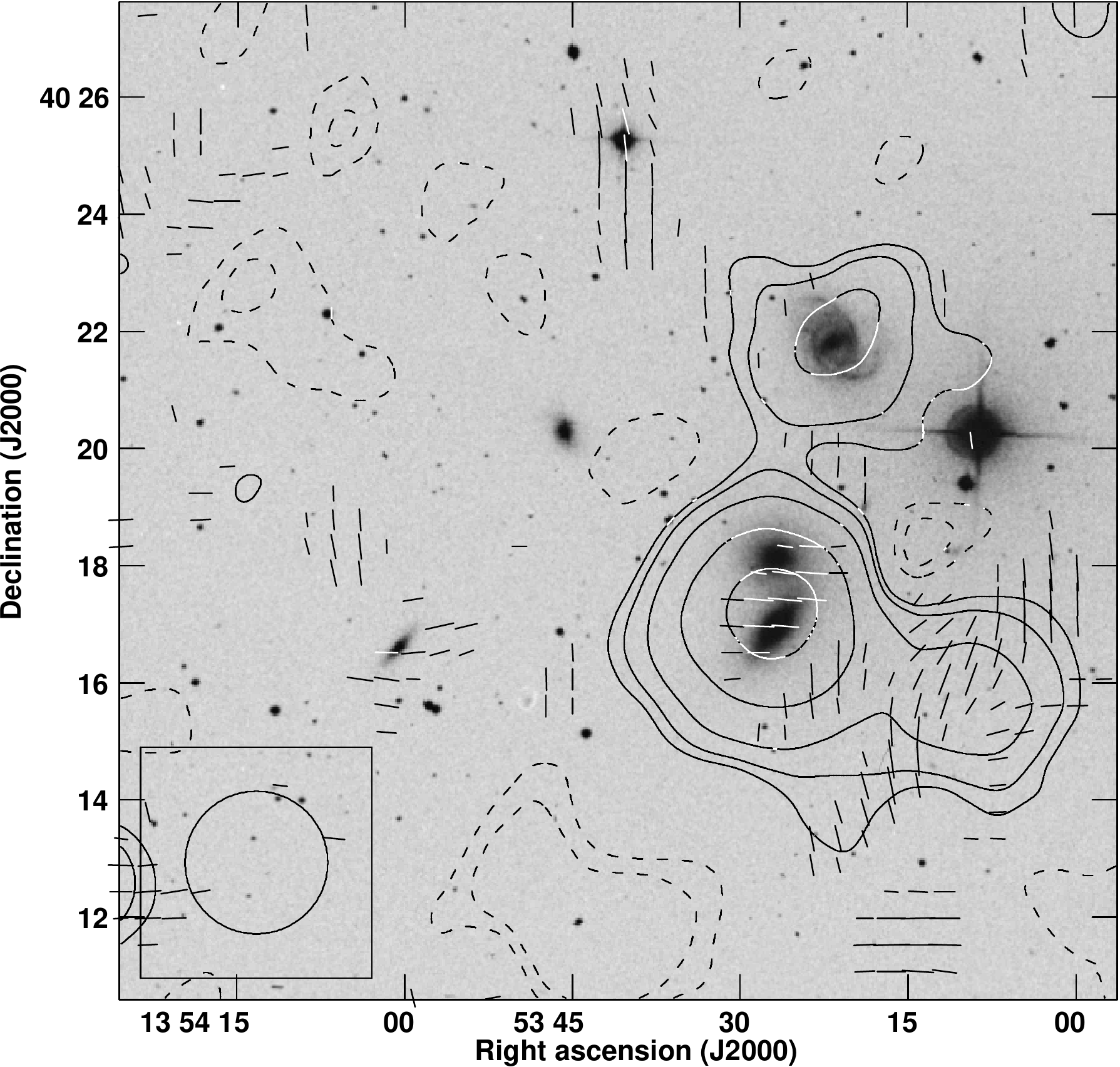}
    \caption{}
    \label{68eff}
    \end{subfigure}%
    \begin{subfigure}[p]{0.33\textwidth}
    \includegraphics[width=\textwidth]{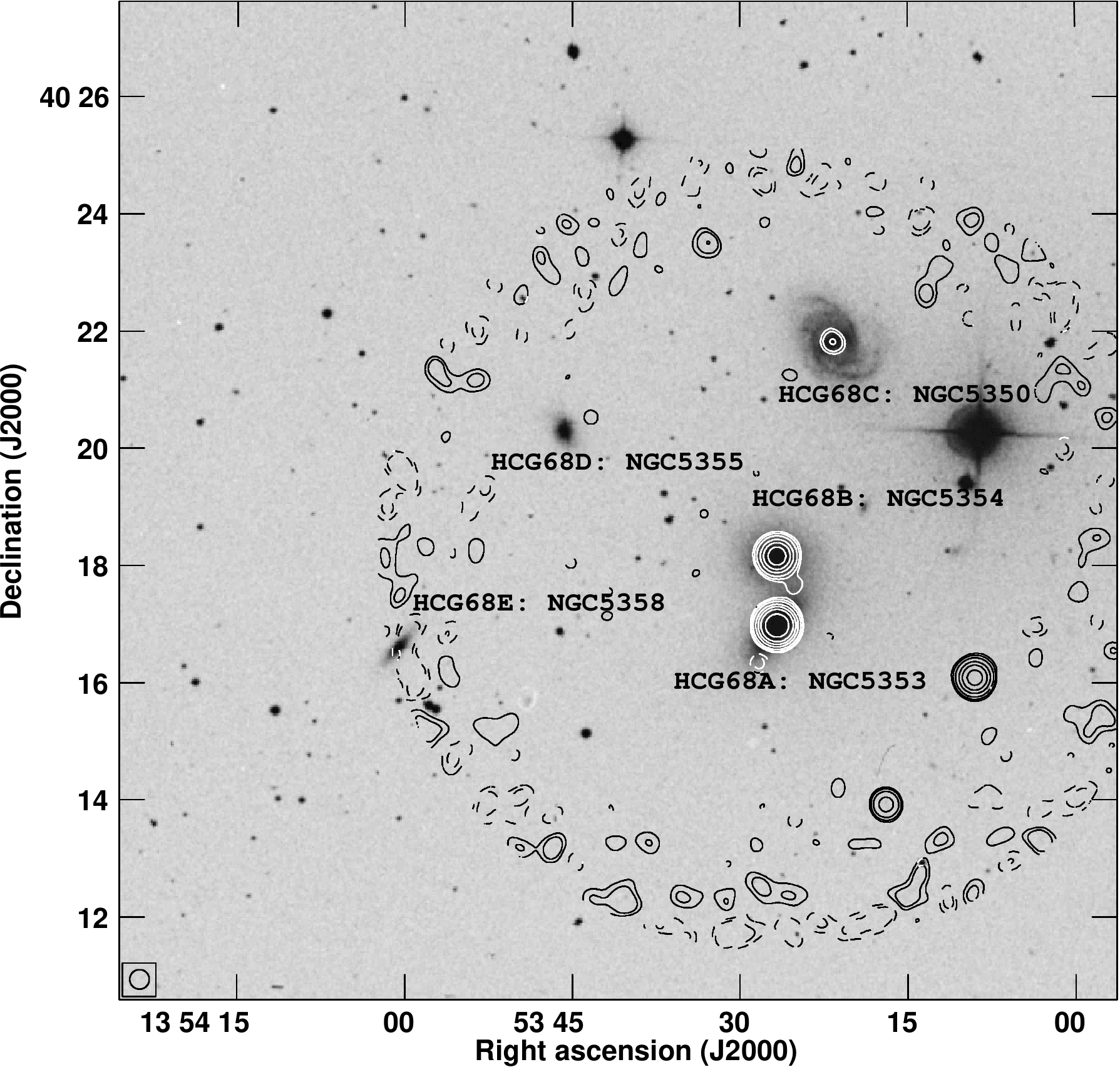}
    \caption{}
    \label{68cnvas20}
    \end{subfigure}%
    \begin{subfigure}[p]{0.33\textwidth}
    \includegraphics[width=\textwidth]{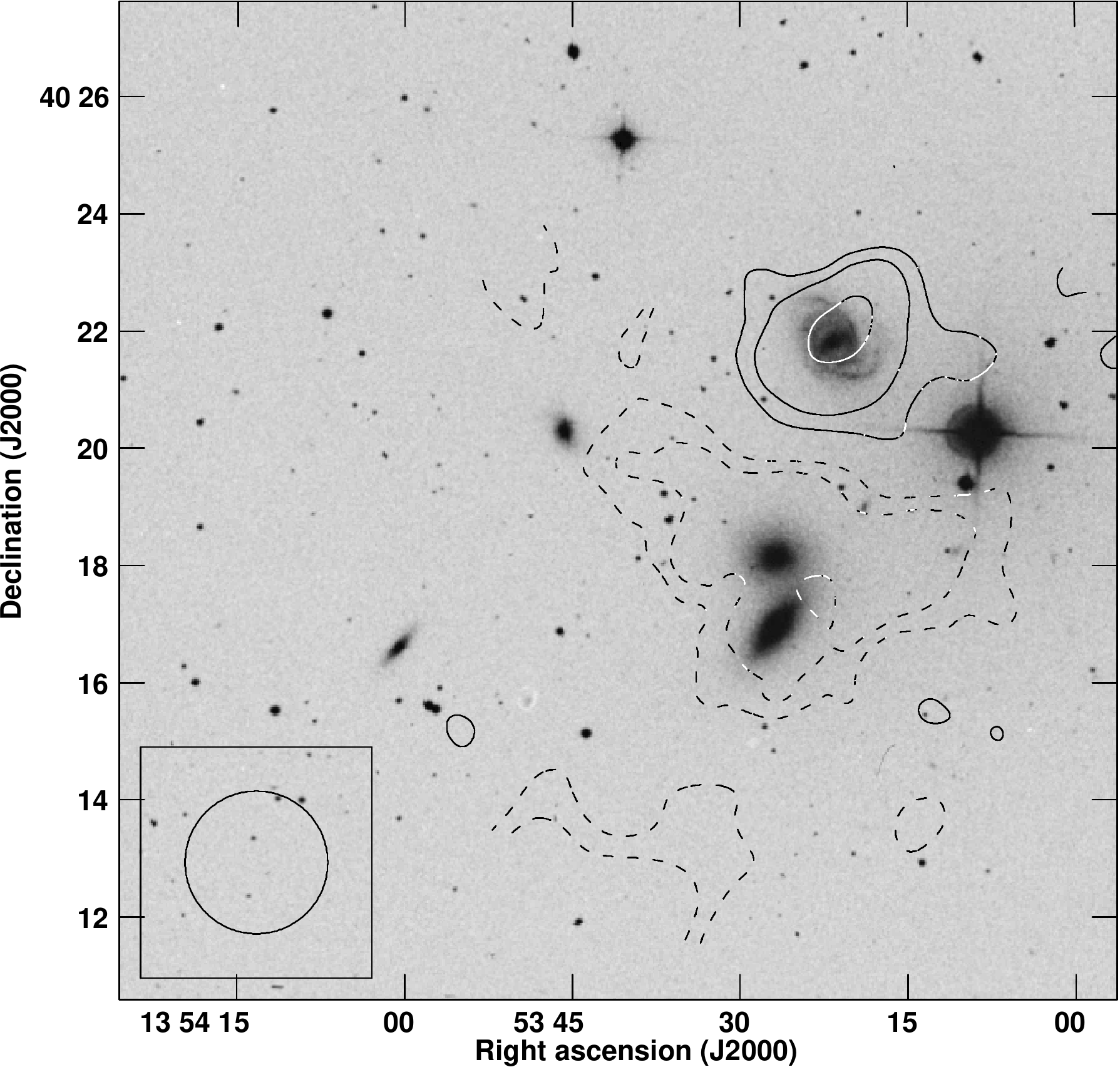}
    \captionsetup{width=0.9\textwidth}
    \caption{}
    \label{68esub}
    \end{subfigure}%

    \begin{subfigure}[p]{0.33\textwidth}
    \includegraphics[width=\textwidth]{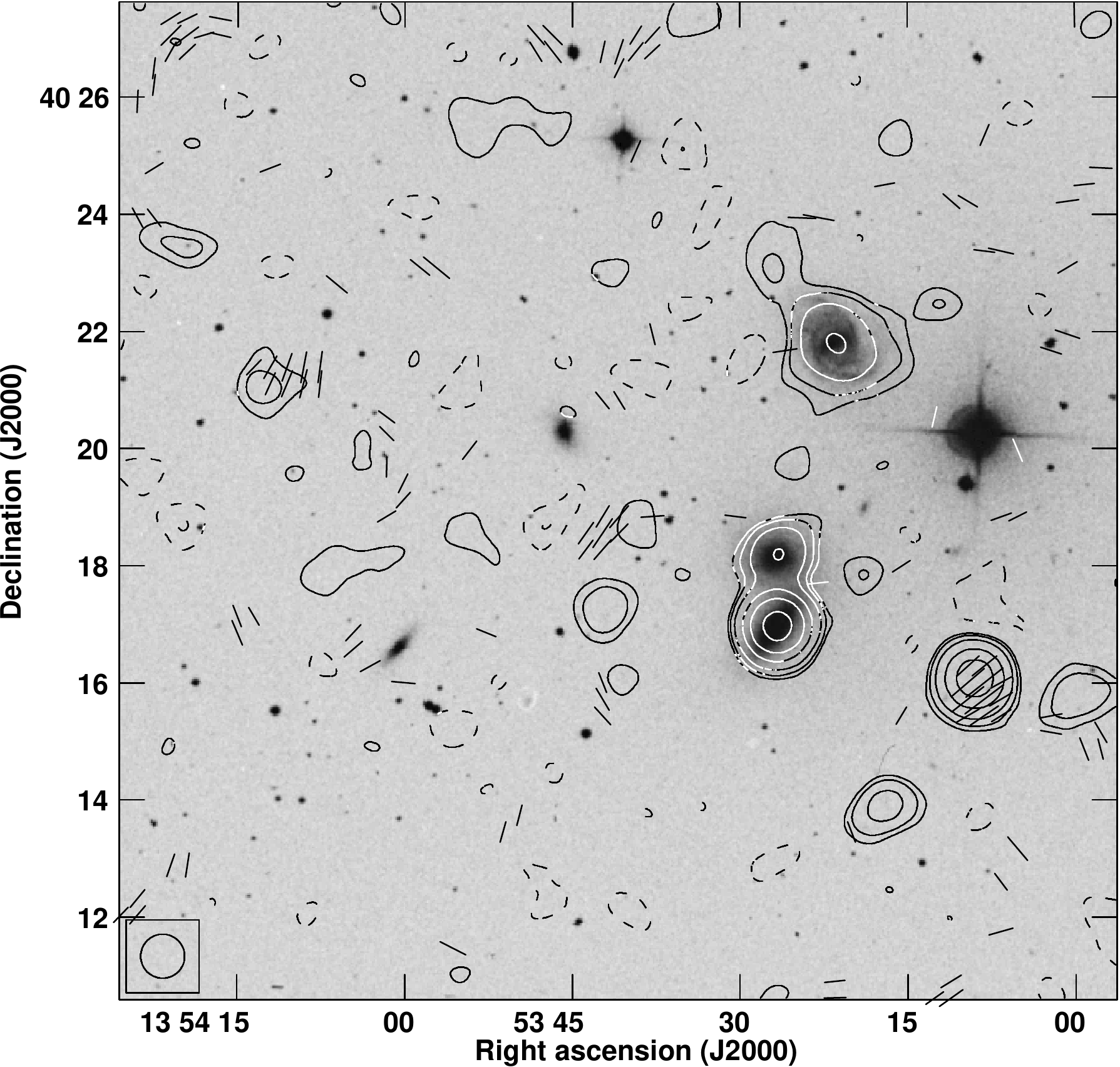}
    \captionsetup{width=0.9\textwidth}
    \caption{}
    \label{68nvss}
    \end{subfigure}%
    \begin{subfigure}[p]{0.33\textwidth}
    \includegraphics[width=\textwidth]{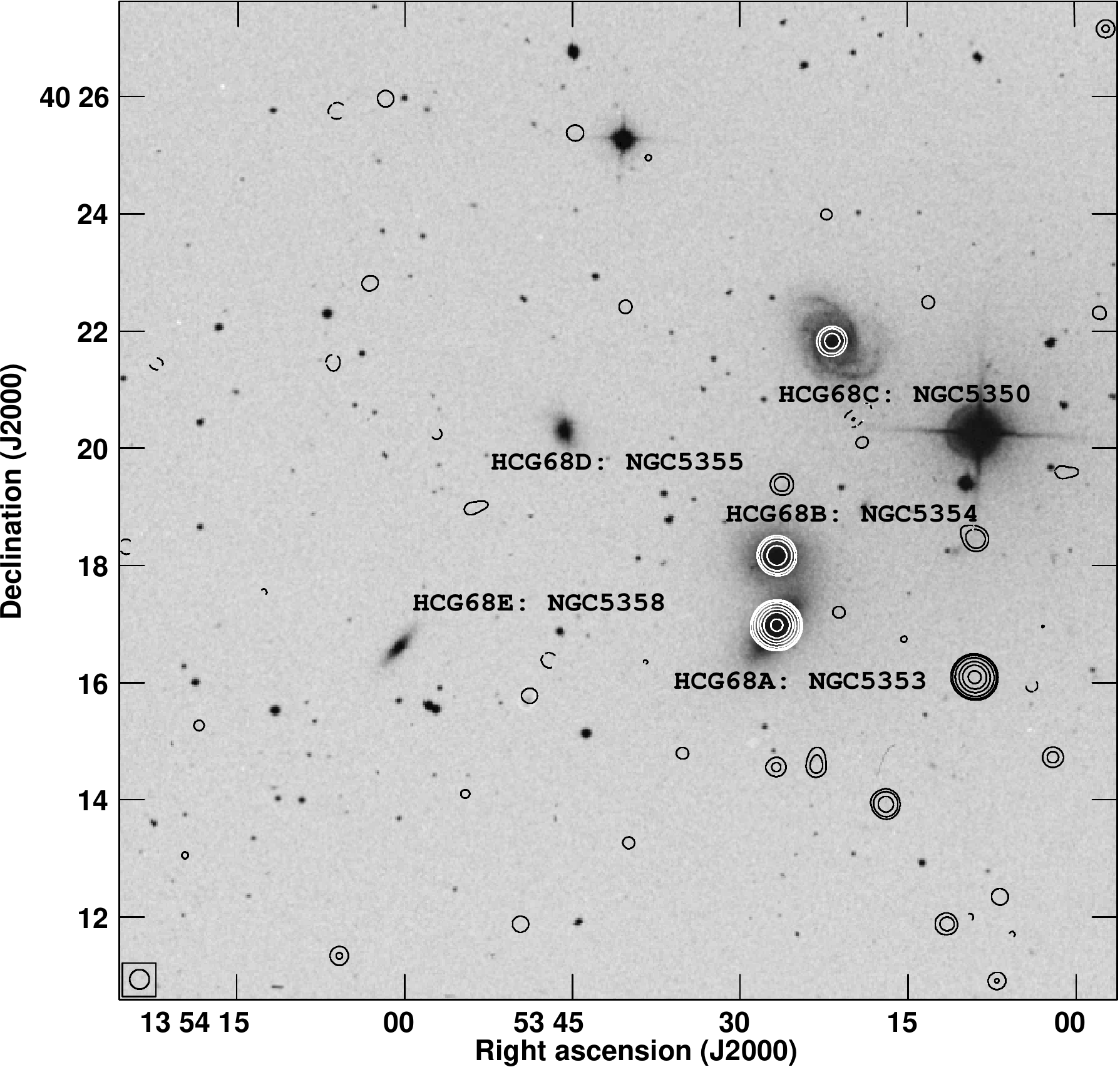}
    \caption{}
    \label{68lnvas20}
    \end{subfigure}%
    \begin{subfigure}[p]{0.335\textwidth}
    \includegraphics[width=\textwidth]{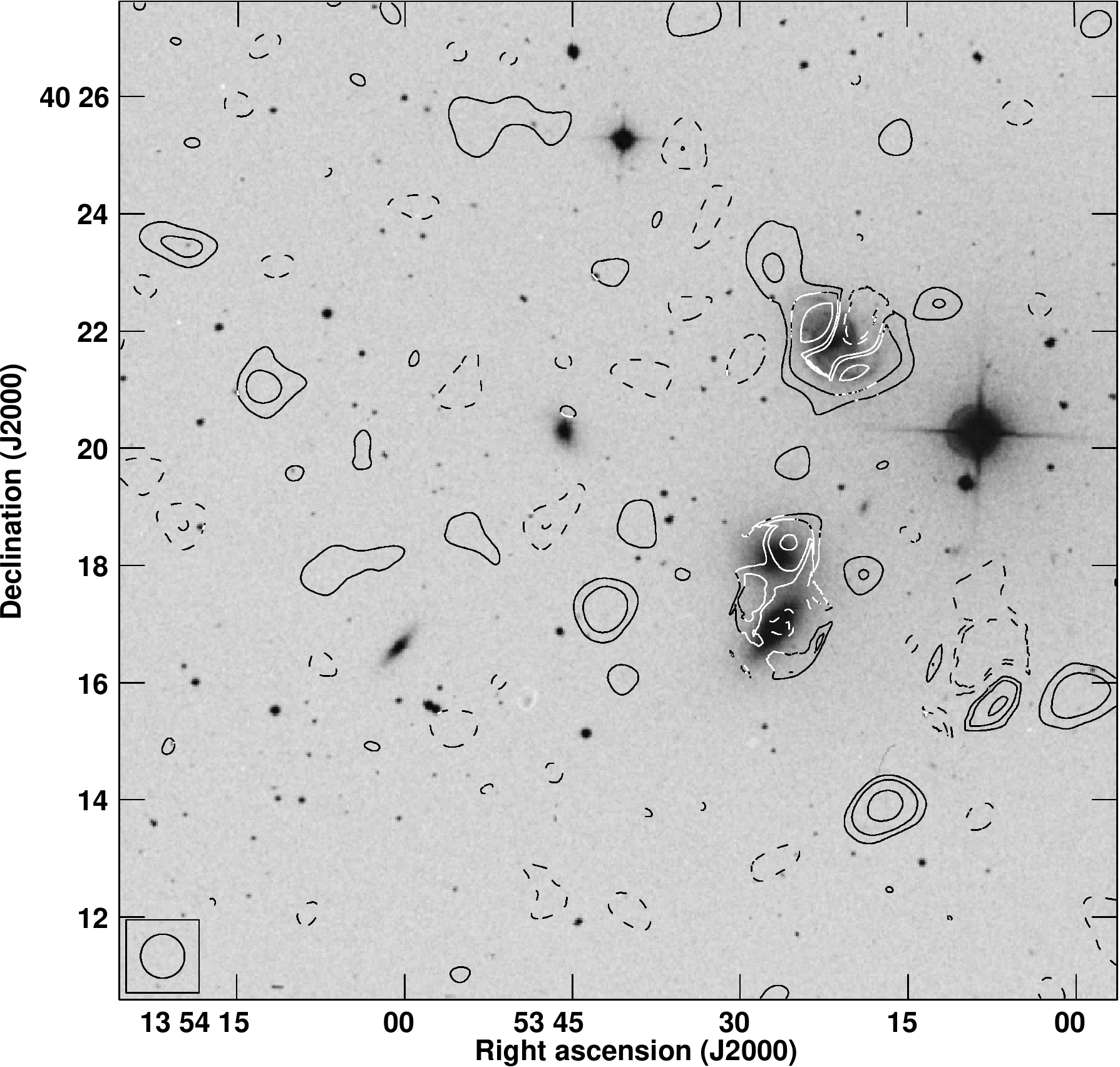}
    \captionsetup{width=0.9\textwidth}
    \caption{}
      \label{68nvsub}
     \end{subfigure}%
    \caption{
          Radio maps of HCG\,68. \textbf{Upper panel:} Effelsberg maps at 4.85\,GHz are shown. 
          \textbf{Lower panel:} NVSS maps at 1.40\,GHz are shown. \textbf{Common for both frequencies:}
          The left map is the TP emission with apparent B vectors overlaid.
          The central map is the TP emission from the NVAS data smoothed to 20 arcsec resolution.
          The right map is the TP emission with compact sources subtracted.
          The background map is a POSS-II R-band image.
          The background map is a POSS-II R-band image.
          The contour levels are $-5,-3$(dashed), $3,5,10,25,50,100 \times$ r.m.s. noise level. 
          The beam is represented by a circle in the lower left corner of the image.
          The 1-arcsec length of the apparent B vectors corresponds to 0.03 mJy/beam;
          1 arcminute is equal to $\approx$ 11\,kpc at the position of HCG\,68.
        }
    \label{fig68}
\end{sidewaysfigure*}

 The final images for the specific galaxies (cropped so only the 
galaxy groups are present, without large void areas) are presented in the right panels of
Figures~\ref{fig15}--\ref{fig68}; 4.85\,GHz results are always in the upper corner and the 1.40\,GHz are in the lower
corner. Flux densities before and after subtraction, as well as the total flux density for all the subtracted sources, are given in Table~\ref{values}. The \textit{subtracted} spectral index value
was calculated using the flux density values measured in the diffuse emission-only maps. As these values are sometimes close 
to the map r.m.s. noise, their uncertainties are much higher. Therefore, they should be treated with 
caution. In several cases when the flux density was given only as an upper limit (r.m.s. value), the flattest possible
index value was inserted. All such examples are listed in the table.\\

It should be indicated here that no polarisation has been detected in the high-resolution
data, rendering the subtraction of the compact, polarised structures impossible.

\section{Results \label{sec:results}} 

Figures~\ref{fig15}--\ref{fig68} consist of six panels. The upper left panel is comprised of contours from the 4.85\,GHz total 
power data with apparent B-vectors of the magnetic field superimposed; the upper middle has contours from the 4.85\,GHz high-resolution
total power data with galaxy designations superimposed; the upper right has contours from the 4.85\,GHz total power data after subtraction
of the compact sources; the lower left has contours from the 1.40\,GHz total power data with apparent B vectors of the magnetic field 
superimposed; the lower middle is comprised of contours from the 1.40\,GHz high-resolution total power data with galaxy designations
superimposed; and the lower right has contours from the 1.40\,GHz total power data after subtraction of the compact sources.\\
All presented magnetic field vectors are apparent, i.e. they have not been corrected for the foreground
Faraday rotation. Table~\ref{values} lists the averaged foreground rotation measure taken from \citet{RM}
and the angle corrections -- upper limits for the foreground Faraday rotation -- are given in the appropriate
paragraphs. Throughout the paper, the $S_\nu \propto \nu^{-\alpha}$ convention is adopted. The $H_{0}$ parameter is assumed
to be 73 km/sec/Mpc.
 
\subsection{HCG\,15}
\label{res-15}
 
The 4.85\,GHz emission of HCG\,15 (Fig.~\ref{15eff}) is dominated by the central engine of
its ``D'' galaxy UGC\,1618. The emission region covers two other galaxies;
however, its extent is similar to the size of the telescope beam. No emission 
has been detected in the Effelsberg observations either from other group 
members or from intergalactic areas. The NVSS map shows a lot more extended
region of emission, covering a vast area of the group. The central galaxy is
still the strongest source of radiation and is embedded in the aforementioned
structure.\\

The polarimetric information in the case of this group is, however, unreliable, as
the signal barely exceeds 3 $\times$ the r.m.s. value.
The NVSS map shows polarisation vectors extending into the intergalactic space
in the central part of the radio halo. However, it is impossible to clarify
whether this is not just a smeared out contribution of background sources.
The vectors are rotated by approximately 90$^{\circ}$ clockwise by
the foreground Faraday rotation.

\subsection{HCG\,44}
\label{res-44}

Unlike HCG\,15 described above, HCG\,44 (Fig.~\ref{44eff}) looks like a much more interesting 
system. Its central galaxy (NGC\,3190) is immersed in a bubble of radio emission
with total flux density of 13.8$\pm$1.0\,mJy. The peak flux density refers to a point
that is located south from the centre of this galaxy. A bridge of radio emission that reaches 
5 $\times$ r.m.s. contour near NGC\,3190, extends northwestward pointing at another member
object, i.e. the barred spiral galaxy NGC\,3187; surprisingly, this bridge is visible only at 4.85\,GHz.
The disk of NGC\,3187 is not visible at
4.85\,GHz. The southernmost member, NGC\,3185, is also surrounded by the radio emission,
and, similar to the case of the central galaxy, the peak flux density of the radio emitting region is 
shifted from its optical centre -- this time, in the northern direction. The elliptical 
NGC\,3193 was not detected at both frequencies.\\

Not much polarisation can be seen from this galaxy group. At 4.85\,GHz, the disk of NGC\,3190 seems
to be polarised. However, at 1.40\,GHz, the polarised signal comes instead from the background 
sources south from it. The foreground Faraday rotation is negligible at 4.85\,GHz, but at the 
lower frequency it alters the apparent B-vector orientation by approximately --30$^{\circ}$
clockwise.

\subsection{HCG\,60}
\label{res-60}

HCG\,60 is dominated by a radio loud region in the centre of this galaxy group. Its 
size is significantly larger than the beam area; the 4.85\,GHz total flux density of 244 $\pm$ 15 mJy makes it
the brightest galaxy group in our study; it is also the brightest group at 1.40\,GHz (657 $\pm$ 33 mJy).
The morphology of radio emission is very similar at both of these frequencies. 
The high-resolution images reveal that this group hosts a head-tail radio source, previously
reported and studied by \citet{rudnick76,miley77}, and \citet{jagers87}.
The 4.85\,GHz map of HCG\,60 has the highest noise level among the set. This is because of a limited
number of coverages. Hence, the polarisation information cannot be used.
Nevertheless, the radio emitting region is easily distinguishable from its surroundings.

\subsection{HCG\,68}
\label{res-68}

The last object in our study is the group HCG\,68. At 4.85\,GHz the radio emitting structure
covers three group members: northernmost NGC\,5350 is connected with the close pair NGC\,5353 and 5354 
through a narrow trail that is most probably an effect of the beam smearing. The whole envelope extends far beyond the closest vicinity of the
member galaxies, also containing an object without an optical counterpart (most possibly
a background source) on its western side. Its northern part forms an envelope
surrounding NGC\,5350. At 1.40\,GHz, there is no connection between this galaxy and the aforementioned
pair, but hints at the presence of an envelope remain. Other group members are radio quiet\ and there are
no intergalactic structures that connect them either at 1.40 or at 4.85\,GHz.\\ 

Several spots of polarised emission can be easily seen at the higher frequency. However, the signal-to-noise ratio for these 
objects is rather low: it barely exceeds 3 $\times$ r.m.s. and in most of the cases is lower than 5 $\times$ the r.m.s.
Therefore it is possible that the emitting area is not a real signal. This claim 
seems to be supported by the results from the 1.40\,GHz emission, where only the western background source is polarised.

\begin{sidewaystable*}
\caption{Flux densities (with uncertainties) obtained for the selected regions of radio emission. }
\begin{center}
\begin{tabular}[]{lcccccccccccc}
Group   & Source        & \multicolumn{4}{c}{Flux density in mJy at a given frequency}      & \multicolumn{2}{c}{$\alpha$}  & B     & E$_{B}$       &RM     \\
No.     &               & \multicolumn{2}{c}{4.85\,GHz} & \multicolumn{2}{c}{1.40\,GHz} & \multicolumn{1}{c}{total} & subt. &   & 10$^{-12}$&   \\
        &               & total              & subt.  & total              & subt.   &                  &         &        $\mu$G         & $\mathrm{erg}\,\mathrm{cm^{-3}}$  & rad/m$^{2}$\\ 
15      & Group         &   5.7 $\pm$  1.3 &  0.9  &   27.3 $\pm$  2.3 &   25.1  & 1.3 $\pm$ 0.5 & > 2.0**        & 4.7 $\pm$ 1.3        & 0.9 $\pm$ 0.5 & +40 \\
44      & N3190         &  13.8 $\pm$  1.0 &  9.2  &   44.0 $\pm$  2.6 &   24.1  & 0.9 $\pm$ 0.2 & 0.8 $\pm$ 0.3 & 6.7 $\pm$ 1.3         & 1.8 $\pm$ 0.7 & -65 \\
44      & N3185         &   4.0 $\pm$  1.0 &  ***  &    7.0 $\pm$  1.1 &   7.01  & 0.5 $\pm$ 0.6 & ***            & 6.0 $\pm$ 1.5                & 1.5 $\pm$ 0.7 & -65 \\
44      & N3187         &   4.4 $\pm$  0.7 &  2.8  &    6.5 $\pm$  0.8 &   5.14  & 0.3 $\pm$ 0.4 & 0.5 $\pm$ 0.6 & 6.9 $\pm$ 1.5                 & 2.0 $\pm$ 0.8 & -65 \\
60      & Group         &   244 $\pm$   15 & 73.2  &    657 $\pm$   33 &    356  & 0.8 $\pm$ 0.2 & 1.3 $\pm$ 0.4 & 3.2 $\pm$ 1.7                 & 0.5 $\pm$ 0.4 & +15 \\
68      & N5350         &   6.5 $\pm$  0.6 &  5.9  &   21.0 $\pm$  1.4 &   19.6  & 1.0 $\pm$ 0.2 & 1.0 $\pm$ 0.4 & 4.6 $\pm$ 2.0                 & 1.0 $\pm$ 0.7 & +10 \\
68      & N5353/54      &  33.9 $\pm$  2.0 &  0.5* &   56.8 $\pm$  3.5 &   17.0  & 0.4 $\pm$ 0.2 & > 2.0**        & 6.7 $\pm$ 1.8                & 1.9 $\pm$ 1.0 & +10 \\
\hline
\hline
\end{tabular}
\end{center}
\begin{flushleft}
$^{*}$~r.m.s. noise value\\
$^{**}$~lower limit (the spectrum is not flatter than given value)\\
$^{***}$~out of the primary beam area of the high-resolution data\\
\textbf{Column designations, from left to right:}
 1 -- Group number;
 2 -- Source name/description;
 3 -- Integrated 4.85\,GHz flux density with uncertainty, expressed in mJy;
 4 -- Integrated 4.85\,GHz flux density after subtraction of compact sources, expressed in mJy (4.85\,GHz flux density excess);
 5 -- Integrated 1.40\,GHz flux density with uncertainty, expressed in mJy;
 6 -- Integrated 1.40\,GHz flux density after subtraction of compact sources, expressed in mJy (1.40\,GHz flux density excess);
 7 -- Spectral index between 1.40 and 4.85\,GHz with uncertainty;
 8 -- Spectral index between 1.40 and 4.85\,GHz after subtraction of compact sources, with uncertainty;
 9 -- Total magnetic field strength with uncertainty, expressed in $\mu$G;
10 -- Magnetic field energy density with uncertainty, expressed in 10$^{-12}\mathrm{erg}\,\mathrm{cm^{-3}}$;
11 -- Approximate foreground Faraday Rotation Measure based on \citet{RM}, expressed in rad/m$^{2}$.

\end{flushleft}
\label{values}
\end{sidewaystable*}

\section{Discussion 
\label{sec:discussion}} 

\subsection{Magnetic field estimation}
\label{magfield}

The magnetic field was estimated in specific areas of the studied objects following the 
formulae derived by \cite{BFELD}. An assumption of (revised) energy equipartition between
the magnetic field and cosmic ray particles was made. All the estimations were made via the \textsc{bfeld}
code presented by \cite{BFELD}. This code derives magnetic field parameters based on the supplied
values of the proton--to--electron energy density ratio $K_0$, total
path length through the source $D$, the mean synchrotron surface brightness of the chosen region,
and the non-thermal spectral index $\alpha$. The value $K_0$ was fixed at the typical value of 100.
We caution that in general this parameter may adopt other values and its
proper estimation is difficult; however, the magnetic field strength is only weakly dependent on it.
The spectral index was calculated between the NVSS and Effelsberg maps after subtraction of the
compact sources; it complies with the  $S_\nu \propto \nu^{-\alpha}$ convention. As the thermal fraction of the
emission is unknown, we assumed that the synchrotron spectral index required in magnetic field estimation can be well 
approximated by our calculated \textit{subtracted} value. In case of objects where the \textit{subtracted} value was unreliable because they were not 
detected in the subtracted 4.85\,GHz maps, the \textit{total} value was adopted (see specific subsections for details). The 
path length was calculated one of two ways. For groups no. 15 and 60 it was calculated under the assumption 
of spherical symmetry of the studied object; however, this probably gives an upper path-length limit so that 
both strength and energy density are higher than those estimated. In case of groups no. 44 and 68, the path length was calculated assuming the magnetic 
disk thickness of 1--2\,kpc; this presumption is widely used in literature (see e.g. \citealt{beck96rev} and references therein). The total uncertainty of the estimated values 
accounts for a factor of two uncertainties rising from the unknown thermal fraction of the emission, 
exact path length, and surface brightness. All the estimated values can be found in 
Table~\ref{values}. 

\subsection{Overview of the individual groups}

\subsubsection{HCG\,15}
\label{dis-15}

This group was expected to be the most 
promising target of our study. However, as already mentioned in Sect~\ref{res-15},
the extent of the radio emission from this 
galaxy group is comparable to the size of the Effelsberg telescope beam at 4.85\,GHz. Subtraction
of the compact sources, in this case, the AGN in the central galaxy UGC\,1624, leaves nothing but noise,
as less then 10\% of the original flux density remains.
There are no signs of over-subtraction, as the negative signal barely extends the 3 $\times$
r.m.s. contour only in isolated, small areas. 
HCG\,15 is known to have its galaxies shrouded in a large halo and is visible very well both 
at 612\,MHz and in the X-ray regime \citep{giacintucci11}; a large part of this structure
is still detectable at 1.40\,GHz in the NVSS and more than 80\% of the total flux density at that
frequency is attributed to the diffuse structure. Even more surprising is the value of the 
spectral index; assuming the noise level as the flux density at 4.85\,GHz and using the NVSS 
signal with the AGN contribution subtracted, one arrives at $\alpha$ of $\approx$ 2.7. 
The unusual steepness cannot be attributed to the incomplete subtraction of
the flux density of point sources at 1.40\,GHz. Even if the maximum subtraction method \citep{chyzy03} is used, the residual
flux density is sufficient to yield a spectral index of $\approx$ 2.2; to achieve a typical index
of $\approx$ 1.0, more than 80\% of total flux density should be subtracted, and this value is clearly too high.
Having in mind that the NVSS is a snapshot survey and the faintest emission can still remain undetected, 
the value of 2.7 can be treated as the flattest spectral index possible.
The unusually steep spectrum is consistent with an age estimate based on the
calculation of time needed for the electrons to travel from the centre of the group to its
outskirts (a distance of app. 55\,kpc): assuming they travel with Alfv\'en speed (here adopted as 500 km/s), 
it takes $\approx$ 110\,Myr to produce the observed
radio halo, which is a reasonably large amount of time to explain the spectral index value.\\

To perform the 
magnetic field estimation, we assumed that the spectral index value is between 2.0 
and 3.0. Spherical symmetry has been adapted to derive the path length; from the angular 
size and distance to the group, values between 15 and 25\,kpc seem to be reasonable.
Substitution of these values to \textsc{bfeld} yielded magnetic field strength of $4.7 \pm 2.3$ $\mu$G
and magnetic field energy density of $9.3 \pm 4.6 \times 10^{-13}$ $\mathrm{erg}\,\mathrm{cm^{-3}}$. These values are
lower than those estimated for already studied compact galaxy systems (such as the Stephan's 
Quintet or Taffy galaxies. In particular, the magnetic field energy density is about a factor of several
lower than for the aforementioned objects. As there was no detection of polarised
signal, it is impossible to estimate the ordered component strength; substitution of noise values to 
derive the upper constraint is also not feasible.\\

If calculated via the 1.40\,GHz flux density and the same (assumed) spectral index, the magnetic field strength 
rises to $7.8 \pm 1.8 \mathrm{\mu G}$. The apparent difference is due to the assumed spectral index value not
accounting for, for example spectrum curvature. The magnetic field energy density derived this way increases to half of the 
value estimated for  Stephan's Quintet and total field strength reaches values typical for galaxies.

\subsubsection{HCG\,44}

At a distance of approximately 20\,Mpc away, HCG\,44 is the nearest object in our study
(and the nearest HCG; \citealt{hickson92}). Its central galaxy NGC\,3190 dominates 
the group at both frequencies. Subtraction of the compact sources, including
two background objects south from the galaxy disk leaves an extended structure that is still significant (nearly 70\% of the original flux density), with the total flux density of $9.2 \pm 1.0$ at 4.85\,GHz. The strength of the 
magnetic field  in the disk of NGC\,3190 is thereby $6.7 \pm 1.3$ $\mu$G and its energy density is equal to 
$1.8 \pm 0.7 \times 10^{-12}$ $\mathrm{erg}\,\mathrm{cm^{-3}}$. Such values are comparable to what
was found for most of the normal (non-starburst and non-AGN) spiral galaxies by \citet{niklas95},
i.e. $9 \pm 1.3$ $\mu$G. The emission seems to be polarised with apparent polarisation
vectors visible over the galactic disk. 
For the southern galaxy, similar values of B$_{\rm{TOT}}$ and E$_{\rm{B}}$
have been derived and therefore similar conclusions can be drawn.\\

HCG\,44 is known to contain a large HI tail \citep{serra13}, i.e. 300\,kpc long. This tail is possibly
a remnant of the encounter between NGC\,3187 and other group members. Albeit the tail itself is located 
north of the group centre, it is still interesting to search for any radio structures that 
might result from this interaction. There is one intergalactic structure that draws
attention: a bridge-like extension between NGC\,3190 and 3187. If real, it would suggest 
that the interaction--driven stripping of the neutral and ionised matter proceeded in a 
significantly different way. However, the detected structure is relatively weak, just exceeding the 3 $\times$ r.m.s. level.
Its angular extent, compared to the beam size, is small. Having in mind that it lies close
to a strong source (disk of the central galaxy and background sources) it is possible that it could be a
scanning effect. Although careful analysis of individual projections failed to identify any single maps
that could cause emergence of the bridge, the limited parallactic angle coverage certainly makes performance of the 
basket weaving method worse. Despite having an angular size that is smaller than the size of the largest observable structure at 1.40\,GHz,
there is no trace of its counterpart in the NVSS data; however, on the other hand, the NVSS consists of snapshots and 
may miss a significant amount of faint, extended emission. Additional coverages and additional data 
from other telescopes are necessary to confirm or reject its existence.\\

Similarly to the case of HCG\,15, we tried to estimate the age of the bridge. If electron velocity equal to the
Alfv\'en speed (adopted as 500\,km/s) is assumed, then the age the structure is equal to approximately 25 Myrs. A supposedly flat spectrum 
is coherent with this value.

\subsubsection{HCG\,60}
\label{dis-60}

HCG\,60 is one of those HCGs that actually do not meet the criteria to be identified
as compact groups. Because it is in the central region of the rich cluster Abell\,1452, HCG\,60 does
not fulfil the isolation requirement. Nevertheless, it is still an interesting target,
confirming the claim that physical conditions in some of the compact groups might be 
similar to those in the central regions of galaxy clusters \citep{hickson82}. The central galaxy of HCG\,60 
has an active nucleus that was originally described by \citet{rudnick76}. \citet{miley77} suggested that 
the unique morphology of this object might be due to a sharp change in the propagation 
direction of the host galaxy through the intra-group medium, and (or) its immersion in the intergalactic wind. 
\citet{jagers87} compiled their results together with his new data at 0.6\,GHz,
concluding that the first possibility is supported by his findings. A study of the 
polarised emission from this object was also presented.\\

Compared to the J\"agers maps, the NVSS data seem to show more polarised emission. However, both
at 4.85, and 1.40\,GHz instrumental effects seem to dominate the picture, yielding the analysis 
of ordered magnetic field component impossible.
Subtraction of the compact sources leaves a significant remaining structure both at 1.40 and 
4.85\, GHz (approximately 55\% and 30\%, respectively). Assuming spherical symmetry 
(thus, path length of 100--300\,kpc), we estimate the magnetic field to be equal
to $3.2 \pm 1.7$ $\mu$G.\\

It is interesting to ask what is the origin of the emission that remains after subtraction of the
compact sources. Both 1.40 and 4.85\,GHz high-resolution, archive data reproduce the head-tail structure of the
radio galaxy very well, so all the structures resulting from recent activity could be efficiently
subtracted. This would leave the remaining emission to be either emerging from old lobes, from
non-AGN-related intergalactic structures, or from both. This claim is supported by the 
comparison of ``extended'' and ``compact'' emission fractions at both frequencies. At 1.40\,GHz,
the ``extended'' component dominates over the ``compact'' component, whereas this situation gets 
reversed at 4.85\,GHz, where the flux density losses cannot be attributed to the missing flux density problem.
However, at this moment it is impossible to exclude any of the presented, possible origins of that 
``extended'' component.\\

\subsubsection{HCG\,68}

The last object in our study is HCG\,68. This group is very close to the Milky Way 
($\approx$ 30\,Mpc) and its main galaxies are listed under the NGC designation. The most recognisable
part of the group is the triplet of galaxies NGC\,5350--5353--5354. All of these galaxies are 
radio loud, with NGC\,5350 classified as either LINER, or starburst \citep{chapelon99} and NGC\,5354 as a possible 
Seyfert galaxy \citep{veron-cetty06}. Subtraction of the compact sources at 1.40\,GHz leaves most of the 
emission from NGC\,5354 intact, as well as two small extensions of the envelope of NGC\,5350. At 4.85\,GHz,
the radio emission from two central galaxies and the nearby background source is significantly 
oversubtracted, with negative signal exceeding the 5x r.m.s. level in an area about one beam in size.
The reasons for this effect are yet unknown; possible explanations include the variability of
the AGN during the period between the high-resolution observations made with the VLA (more than 25
years), or that additional flux density errors induced by the weather conditions and not completely ruled
out by the beam-switching system; hence, re-observations for this group might be of use.
Therefore, the total (non-subtracted) spectral index was used to estimate the magnetic field properties of 
NGC\,5353/54. The resulting values suggest a rather typical field of microgauss strength.
The northernmost galaxy NGC\,5350 possesses a slightly weaker field. Both at 1.40 and 4.85\,GHz, the size
of the radio halo is larger than that of the optical disk. However, there are no signs of outflows from the
halo or its distortions, which could signify ongoing interactions.

\subsection{Question of the intergalactic magnetic fields}

The main reason behind this study was to estimate the intergalactic 
field strength and energy density for a selected sample of HCGs in a manner particularly sensitive to weak extended 
radio emission. This study combined with our previous investigations 
of intergalactic emission from galaxy groups has revealed the following 
(phenomenological) classes of such objects: 

\begin{itemize}

\item[1.] Spiral-dominated groups with no detectable intergalactic 
emission: Leo Triplet \citep{bnw13A}, and HCG\,68 (possibly also HCG\,44)
\item[2.] Spiral-dominated groups with strong intergalactic magnetic 
fields: Stephan's Quintet (SQ, \citealt{bnw13B})
\item[3.] E/S0-dominated groups with intergalactic emission and a
currently active nucleus in at least one of the member galaxies: HCG\,60
\item[4.] E/S0-dominated groups with intergalactic emission but no 
jet-producing nuclei: HCG\,15

\end{itemize}

In HCG\,68, the only large-scale  structure could be the halo 
of its member galaxy NGC\,5350. To make any conclusions much 
better data are needed; nevertheless, it is likely
that there is no genuine diffuse intergalactic radio emission
in this system.\\ 

What could generally be expected is that systems of galaxies that
usually host magnetic fields would be those that contain intergalactic
magnetic fields as well.
Late-type spiral galaxies are known to be sources of magnetic 
fields: phenomena that are responsible for their effective amplification, 
such as the dynamo process, star formation, or supernovae, 
are  common in such objects.  However, the Stephan's Quintet,
(SQ), possessing 
intergalactic radio emission (class (2.) described above) is a much 
tighter group than those belonging to class (1.). Moreover, it shows  
indications of star formation and intergalactic shock between member 
galaxies \citep{xu05}. Indeed the intergalactic emission in SQ has a 
relatively flat spectrum, as it is easily recognisable at frequencies as 
high as 8.35\,GHz \citep{bnw13B}. This indicates a mutual 
proximity between electron-producing, star-forming regions and those 
producing radio emission.\\

The first class of systems contains these objects, which do not
host intergalactic radio emission.
For loose groups, the spatial scale of galaxy distribution may be too 
large to be efficiently filled with magnetic fields and cosmic rays 
pulled out from member galaxies. The differences between magnetic 
fields in loose (Leo Triplet) and tight (SQ) groups are emphasised by 
the fact that the tidal dwarf formed in the gaseous tail is unmagnetised 
in the first case, while housing strong magnetic 
fields ($\ge$ 6$\mu$G) in case of the SQ tidal dwarf \citep{bnw13B, bnw14B}.\\

HCG\,44 falls into one of the two categories described in the
former paragraphs. The weak bridge is the single sign of 
intergalactic magnetic fields, but its low signal-to-noise ratio and lack of
detection at any other frequency makes it hard to discuss.
If this structure is not a real one, it falls into the first category
-- and indeed the size of the group and some of the features (e.g. the 
giant neutral gas tail) bears resemblance with the Leo Triplet; but even if 
the structure is real, still, the extent of the intergalactic emission
and its character are incomparable to the opulence of magnetised 
structures seen in the SQ. HCG\,44 could be regarded as a transitional
state between classes (1.) and (2.), but it does not violate the general 
conclusion that loose groups are less prone to be brimful of  magnetised
IGM.\\

Elliptical, or lenticular, early-type galaxies are usually believed to
lack mechanisms such as the 
large-scale dynamo generating disk magnetic fields or star formation (thus 
supernovae) that produce cosmic-ray electrons. They are not expected 
to contribute to intergalactic radio emission. An 
alternative is the supply of the magnetic fields and relativistic particles via 
jets or large-scale structures propelled by AGNs.  This may be the 
case for HCG\, 60 representing class (3.) as defined above. These kinds of 
radio structures are frequent in the survey by \citet{giacintucci11}. The diffuse 
radio emission seems to form a cocoon encompassing the pair of jets 
and extending even further into extragalactic space. As expected 
for an ageing population of relativistic electrons, 
this radio envelope has a rather (but not extremely) steep spectrum 
(Tab.~\ref{values}). In this class of galaxy groups the intergalactic magnetic 
fields may be spread around by AGNs.\\

The jets of HCG\,60 are perpendicular to each other, which is a clear 
indication of ram pressure caused by galaxy motion through the  
intragroup matter. Indeed, many groups from the HCG list exhibit a 
substantial $\ion{H}{i}$ deficiency (e.g. \citealt{hutchmeier97}) possibly
due to ram pressure effects (or, at least partly). Also,  the estimated 
intergalactic magnetic fields in the cocoon  of HCG\,60 are rather weak. 
However, as mentioned in Sect.~\ref{dis-60}, the path length through 
the emitting matter was estimated assuming a spherical symmetry of 
the cocoon. Thus, the obtained field strength and energy density can 
be regarded as lower limits while the true values can be somewhat 
higher.\\

HCG\, 15 represents the most interesting objects denoted above as 
class (4.). Its early-type member galaxies lack star formation signatures 
(hence dynamo mechanism and sources of relativistic electrons). They 
also lack strong, jet-producing AGNs, the origin of intergalactic 
synchrotron emission is therefore not obvious. \citet{darocha08} found a faint 
stellar glow in the intragroup space of HCG\,15; however, its colour 
resembles that of the old stars. Such stars cannot produce supernovae,
so the cosmic rays might originate only from supernovae remnants from former
generation of stars; this old electron content is unlikely to explain the 
observed radio structure.
\citet{giacintucci11}  suggested that the radio structure of HCG\,15 may be 
an analogon of the shock region in Stephan's Quintet. Apparently 
this is not case because of the lack of traces of intergalactic 
starburst activity \citep{darocha08}.\\

The lack of identifiable sources of  intra-group magnetic fields in 
HCG\,15, together with its very steep high-frequency spectrum (Tab.~
\ref{values}), bears some similarities with halo-like relics found in clusters. 
\citet{slee01}, following \citealt{giovannini99}, states that approximately 10\% of 
the clusters contain such structures. Even though most of the relics are arc-like objects, 
nevertheless there are also a number of halo-like relics, such as the relic in the
Coma cluster \citep{willson70}. 
The sizes of such structures, usually 1--2\,Mpc, 
are significantly larger than the relic found in  HCG\,15, however clusters 
themselves are also larger than galaxy groups; in addition, the so-called
mini-haloes of hundreds of kpc have also been reported. With its spatial extent
of $\approx$ 100\,kpc, the radio envelope of HCG\,15 can possibly be treated as
a lower boundary for such structures.
\citet{slee01} surveyed several 
cluster relics and found their high-frequency spectral indices to be in 
the range 2.1 -- 4.4, which is considerably steeper than for the large-scale 
Coma halo (1.34, \citealt{kim90}) and  dramatically steeper than for galactic 
disks. HCG\,15 belongs to one of the most $\ion{H}{i}$ deficient 
HCGs \citep{verdes-montenegro07}, which may indicate strong gas-stripping processes, 
hence possible compression phenomena in the intergalactic gas. 
However, polarisation information is still missing in this picture; the
cluster relics are known to be polarised. Unfortunately, without sensitive polarimetric
observations around or below 1\,GHz, the true character of the detected structure cannot
be revealed.

The frequency coverage of our study is too scarce to accurately estimate the age 
of the radio emitting structure in HCG\,15. Nevertheless, a crude 
estimate can be made under the assumption that the break frequency is 
located somewhere between 1.40 and 4.85\,GHz. This is
probable as the flux density values measured by \citet{giacintucci11} yield a total 
spectral index of approximately 0.8 between 0.61 and 1.40\,GHz. The 
contribution of compact sources at 1.40\,GHz is not too high, at about 
10\%, so the low-frequency spectral index of the diffuse emission 
should not be larger than 0.8--1.0, signifying a considerably flatter 
spectrum than at frequencies above 1.4\,GHz. Using the magnetic 
field strength values derived in  Sect.~\ref{magfield} and substituting 
them to the spectral age equation from \citet{murgia96}, we estimate the 
radiative age of the envelope to be between 0.3 and 1.0\,$\times 10^8$ 
years. This is very similar to what \citet{slee01} found for four relics they 
studied, it is also similar to the estimate based on the electron propagation
velocity presented in Sect.~\ref{dis-15}.

Irrespectively of its origin, this intergalactic reservoir of gas hosts 
magnetic  field that, albeit weaker than in individual galaxies, can still 
be dynamically important; its energy density is lower than in Stephan's 
Quintet, but still of the same order of magnitude. The thermal energy 
density of the IGM of HCG\,15 is likely to  be lower than for the 
Quintet if the main source of its heating was the passage of an 
intruder. As the 
electrons have not been supplied for a long time, it must have had taken
place long ago and the gas has 
probably significantly cooled down. Altogether, this suggests that the 
dynamics of the intergalactic gas in HCG\,15 is likely to be  highly 
dependent on the magnetic field contained within it. 

As shown by our study there are more examples of intra-group 
magnetic fields able to contribute pressure to the intergalactic gas 
dynamic. There is clear evidence that the magnetic pressure cannot 
be neglected in modelling the dynamics of  
gaseous processes in the intergalactic gas.

\section{Conclusions \label{sec:conclusions}} 
 
We performed a sensitive search for extended radio emission from 
four galaxy groups from the list of HCGs. We selected the following objects: spiral-dominated 
HCG\,44 and HCG\,68 as well as E/SO-dominated HCG\,15 and HCG\,60.
To ensure the maximum sensitivity for 
extended, diffuse structures we used the Effelsberg 100 m radio 
telescope at 4.85\,GHz. To avoid contamination of extended emission by 
beam-smeared compact structures we used the NVAS
data at the same frequency, convolved to the resolution of the Effelsberg 
telescope, and then subtracted from our observations. To obtain the 
spectral information for our objects we used the archive NVSS data,  
also performing the subtraction of beam-smoothed compact structures 
from high-resolution VLA archive data.\\
\\
The following results were obtained:

\begin{itemize}

\item[-] The spiral-dominated group 
HCG\,68 apparently does not show intergalactic emission, hence the 
intra-group magnetic fields are very weak. The same was found in 
our previous studies for another spiral-dominated group, i.e. the Leo Triplet \citep{bnw13A}. This remains in contrast with our recent finding for 
another spiral-dominated group, which is the much more tightly packed  
Stephan's Quintet, where strong intergalactic magnetic 
fields ($\approx$ 10\,$\mu$G) apparently pulled out from member galaxies were found.
The second spiral-dominated object in the sample, HCG\,44, may
contain a magnetised outflow, but it has too low signal-to-noise ratio 
to be discussed. It might be a transitional case: a still loose object,
but with weak intergalactic emission already present.\\

\item[-] The E/S0-dominated group HCG\,60 contains an 
active radio galaxy with two jets. Their subtraction still leaves a 
diffuse radio emitting envelope with a moderately steep spectrum 
($\alpha$ = 1.3). Apparently the radio galaxy supplies magnetic fields 
and cosmic rays to the intra-group space. Substantial ram pressure
effects, which are probably caused by the intra-group medium, can be seen here.
This phenomenon may also be responsible for the $\ion{H}{i}$ deficiency 
in many groups (e.g. \citealt{hutchmeier97})  \\

\item[-]  The E/S0-
dominated group HCG\,15 has been found to contain extended radio emission, signifying the existence of a
5--7\,$\mu$G magnetic field. This is surprising as this group lacks 
obvious sources of magnetic fields and cosmic rays: the member 
galaxies show no signatures of recent star-forming activity. The colour 
of the optical glow between galaxies \citep{darocha08} implies that only old stars 
are present there. The member galaxies also do not contain jet-forming AGNs. 
Indeed the radio structure shows an extremely steep 
spectrum with $\alpha$ not lower than $\approx$ 2.7, which means that the relativistic 
electron supply ceased long ago. In these respects the diffuse radio 
emission in  HCG\,15 resembles micro-haloes found in galaxy 
clusters (e.g. \citealt{slee01}). What we observe in HCG\,15 may be a scaled down 
in size phenomenon similar to cluster relics; however, polarimetric data 
recorded beyond 1400\,MHz is needed to clarify this hypothesis.

\end{itemize}

Our studies (also the previous ones) show that at least some galaxy 
groups contain large reservoirs of magnetised intergalactic matter. The 
magnetic fields and cosmic rays may be either supplied by star-forming 
spiral galaxies (Stephan's Quintet), by active nuclei in group 
galaxies, or even by processes that bear similarities to those in cluster haloes. Whatever is 
the origin of this magnetism, the magnetic energy density seems 
comparable to the thermal (X-ray) energy density. The role of magnetic forces and pressures 
in the dynamics of the intergalactic gas of such galaxy groups cannot be 
neglected.

\begin{acknowledgements} 
BNW and MS are indebted to the staff of the radio telescope Effelsberg 
for all the help and guidance during the observations.
The authors wish to thank Aritra Basu from the MPIfR Bonn for valuable comments.
We acknowledge the usage of the HyperLeda database (http://leda.univ-lyon1.fr) 
and the NASA/IPAC Extragalactic Database (NED), which is
operated by the Jet Propulsion Laboratory, California Institute of Technology,
under contract with the National Aeronautics and Space Administration.
BNW, MU, and MS acknowledge support from the NCN OPUS UMO-2012/07/B/ST9/04404 funding grant.\\
RB and DJB acknowledge support from the DFG Research Unit FOR1254.
\end{acknowledgements} 
 
{} 
 
\end{document}